\documentclass[prl,aps,floatfix,superscriptaddress,twocolumn]{revtex4-1}
\usepackage{amsmath,amsfonts,amssymb,graphics,graphicx,color,bm}
\usepackage{braket}
\usepackage{bbold}
\usepackage[normalem]{ulem}

\usepackage{natbib}
\newcommand{\vecbf}[1]{{\bf #1}}

\begin{document}

\title{Full Counting Statistics for Interacting Fermions with Determinantal 
       Quantum Monte Carlo Simulations}
\date{\today}
\author{Stephan Humeniuk}
\affiliation{Institute for Theoretical Physics III and Center for Integrated Quantum Science and Technology, University of Stuttgart, 70550 Stuttgart, Germany}
\author{Hans Peter B\"uchler}
\affiliation{Institute for Theoretical Physics III and Center for Integrated Quantum Science and Technology, University of Stuttgart, 70550 Stuttgart, Germany}

\begin{abstract}
We present a method for computing the full probability distribution function 
of quadratic observables 
for the Fermi-Hubbard model within the framework of determinantal quantum 
Monte Carlo.
Especially, in cold atoms experiments with single site resolution, 
such a full counting statistics can be obtained from repeated projective measurements.
We demonstrate, that the full counting statistics can provide important information 
on the size of preformed pairs. Furthermore, we compute the full counting 
statistics of the staggered magnetization in the repulsive Hubbard model at half filling
and  find excellent agreement with recent experimental results. We show that current experiments 
are capable of probing the
difference between the Hubbard model and the limiting Heisenberg model.
\end{abstract}
\maketitle

Full counting statistics has emerged as a very powerful tool to characterize
and obtain information about a quantum mechanical system by 
gaining knowledge on the full probability distribution function of an 
observable rather than just its expectation value. 
These concepts have been pioneered by Lesovik and Levitov  \cite{Levitov1996} 
for transport measurements in nano-structures \cite{QuantumNoise};
a remarkable application being the demonstration of the
fractional charge of quasiparticles in a fractional quantum Hall fluid 
\cite{de-Picciotto1998, Saminadayar1997}.
The concept of full counting statistics turns out to be very powerful in 
the context of cold atomic gases, where 
the observable of interest is the number of particles on a set of lattice sites, 
which is accessible with site-resolved quantum gas microscopes \cite{Bakr2009, Sherson2010}.
Especially, it has been applied to characterize quantum states of cold atomic gases 
in equilibrium \cite{Oettl2005, Cherng2007, Belzig2007, Braungardt2008}  as well as non-equilibrium
states  \cite{Lamacraft2006, Braungardt2011, Gring2012}.

Several cold atoms setups for fermionic atoms are currently equipped with a quantum 
gas microscope \cite{Haller2015, Parsons2016, Cheuk2016, Drewes2016, Mazurenko2017, Hilker2017, Mitra2017}
and have achieved single-site and single-atom detection, 
which is required for the measurement of probability distributions in Fock space 
by accumulating histograms of particle configurations over independent 
measurement realizations \cite{Mazurenko2017}. Notably,
both the repulsive \cite{Parsons2016, Cheuk2016, Drewes2016, Mazurenko2017, Hilker2017}
and the attractive \cite{Mitra2017} Hubbard model have been realized in cold atoms experiments
at temperature scales that may be already relevant in the context of normal-state properties
of high-$T_c$ cuprates.
While for bosonic systems, the full counting statistics is accessible within 
path integral quantum Monte Carlo simulations,
in the fermionic situation the notorious sign problem renders such path 
integral approaches inefficient. In turn, 
for auxiliary field QMC methods such as determinantal quantum Monte Carlo (DQMC) \cite{Assaad2008}, 
suitable to study such fermionic systems in certain parameter regimes, 
there is no direct correspondence between the computational configuration 
space of auxiliary fields and states in Fock space.

In this letter, we demonstrate a method to compute the full counting statistics (FCS) 
within the framework of determinantal quantum Monte Carlo simulations and provide first
comparisons with experiments for fermionic cold atomic gases. The main observation for 
this ability is the fact that DQMC simulations decomposes the interacting
fermionic system into an incoherent sum over density matrices for free fermions in 
an external potential \cite{Grover2013}. For such free fermions, the spectrum of 
the reduced density matrix required
for the determination of the full counting statistics is related to the eigenvalues 
of a one-particle correlation function \cite{Cheong2004,Peschel2003}. 
Furthermore, the generating function for the FCS of the particle number, 
magnetization, and the  staggered magnetization is still quadratic in the fermionic field operators. 
This allows us to calculate the relevant trace over the 
exponentially large Hilbert space of fermionic Fock states in
all particle number sectors as the determinant of 
a single-particle operator.  We demonstrate that full counting 
statistics is a powerful tool for the detection of 
pairing and the size of the pairing wave function
by an odd-even effect in the probability distribution 
for the particle number. Furthermore,
we present the first comparison of the FCS in the 
repulsive Hubbard model at half filling and demonstrate that
the experiments clearly observe signatures 
which are not captured by the Heisenberg model. 

\begin{figure}[b!]
\includegraphics[width=1.0\linewidth]{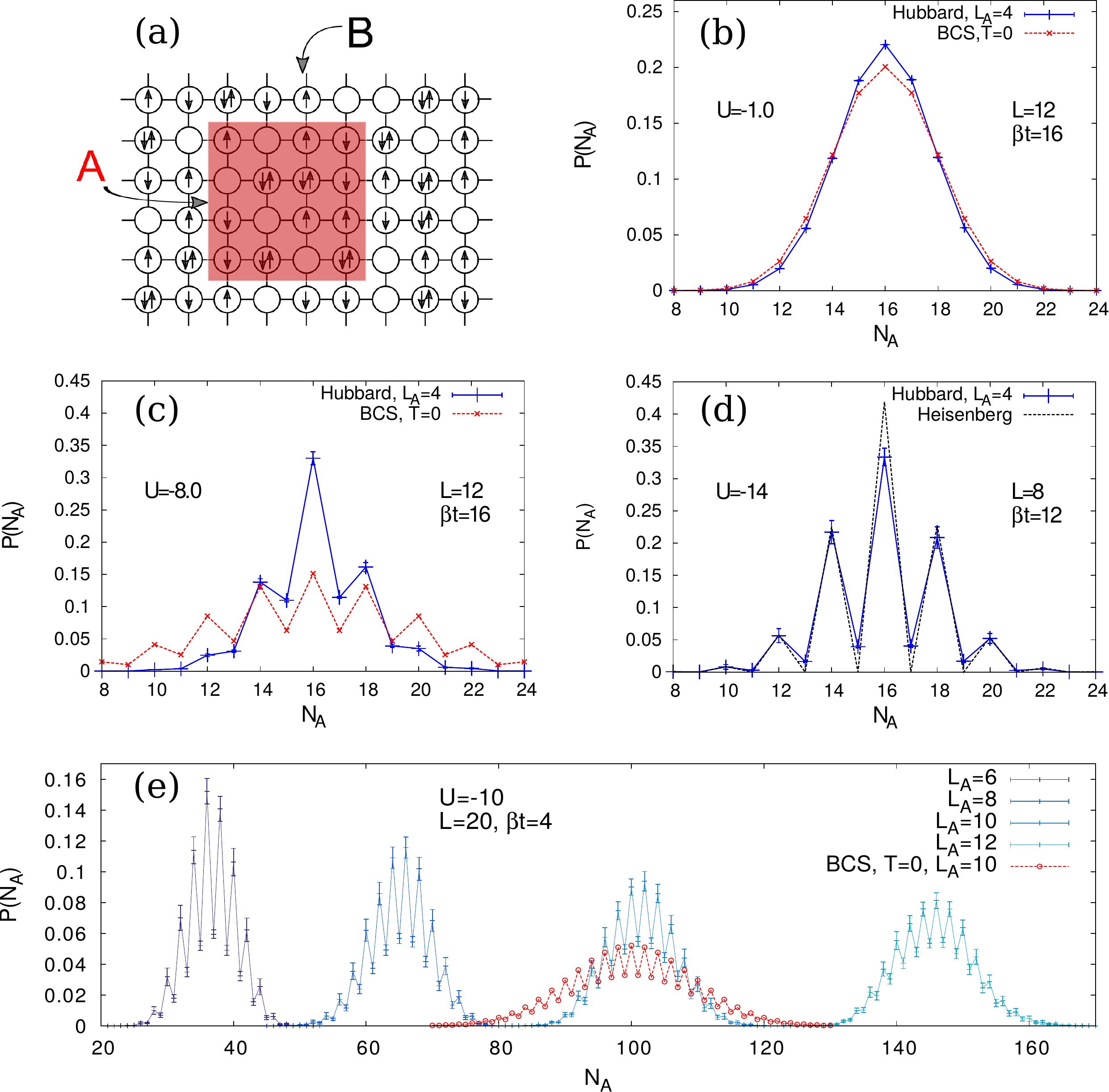}
\caption{(a) Setup of the square subsystem  A  of dimension 
$L_A \times L_A$  embeded in a  larger $L \times L$ square lattice. 
(b-d) Particle number  distribution $P_N(N_A)$  along the 
BCS-to-BEC cross-over at half filling for $U/t=-1,-8,-14$. 
For comparison, the distribution for a BCS mean-field ansatz
at $T=0$ is shown in red, as well as the Heisenberg model for strong interactions.  
(e) Broadening of the distribution function  $P(N_A)$ and 
smearing of the even-odd oscillations for increasing subsystem size $L_A$.}
\label{fig:BEC_BCS_crossover}
\end{figure}

We start with the detailed description of the numerical methods for the determination of the FCS.
The quantity of interest is the probability $P(N^{\uparrow}_{A}, N^{\downarrow}_{A})$ that there are $N^{\uparrow}_{A}$ fermions with spin up and   $N^{\downarrow}_{A}$ 
fermions with spin down on a subsystem $A$ with $N_s$ sites,  see Fig.~\ref{fig:BEC_BCS_crossover}(a). The observable
of interest is therefore given by the spin density operator $\hat{N}^{\sigma}_A = \sum_{i\in A} \hat{c}_{i,\sigma}^{\dagger} \hat{c}_{i,\sigma}$ with 
$\sigma \in \{\uparrow, \downarrow \}$ and eigenvalues $N^{\sigma}_A = 0,\ldots,  N_s$. Its distribution function
is most conveniently derived by the generating function
\begin{equation}
 \chi(\phi^{\uparrow},\phi^{\downarrow}) = \langle e^{i \phi^{\uparrow} \hat{N}^{\uparrow}_A + i \phi^{\downarrow} \hat{N}^{\downarrow}_{A}}\rangle = \text{Tr}\left(\rho_A \, e^{i\sum_{\sigma}\phi^{\sigma} \hat{N}^{\sigma}_A} \right).
 \label{eq:gen_function}
\end{equation}
Here, $\rho_A = \text{Tr}_{B = \bar{A}}\left( \rho \right)$ denotes the reduced density matrix with $B=\bar{A}$ the 
complement of subsystem $A$. Then, the joint probability  $P(N^{\uparrow}_{A},N^{\downarrow}_{A})$ is determined as the coefficient of the Fourier series
of  the generating function 
\begin{equation}
 P(N^{\uparrow}_A,N^{\downarrow}_A) = \sum_{n,m=0}^{N_{s}}  \frac{e^{-i  \phi^{\uparrow}_n N^{\uparrow}_A -i  \phi^{\downarrow}_m N^{\downarrow}_A}}{(N_{s}+1)^2} \chi(\phi^{\uparrow}_n,\phi^{\downarrow}_{m})
 \label{eq:FT_chigen}
\end{equation}
with  $\phi^{\sigma}_n = \frac{2 \pi n}{N_{s}+1}$ . Note that the maximal number of fermions
with a given spin in the subsystem A is limited by $N_{s}$, and therefore, there are only $N_{s}+1$
coefficients in the Fourier series. 
In general, one is mostly interested in the probability distributions  $P_{N}(N_{A})$ 
for the total particle number $N_{A} = N_{A}^{\uparrow} + N_{A}^{\downarrow}$ and $P_{M}(M_A)$ 
for the magnetization $M_A= N_A^{\uparrow} - N_A^{\downarrow}$ on subsystem A. These quantities
derive directly from the joint probability distribution via
\begin{eqnarray}
P_N(N_A) = \sum_{N_A^{\downarrow}=0}^{N_s} P(N_A - N_A^{\downarrow}, N_A^{\downarrow})
\label{eq:PN_marginalization}
\end{eqnarray}
with $N_A = 0,\ldots, 2 N_s$. 
An analogous expression follows for  $P_{M}(M_{A}) =  \sum_{N_A^{\downarrow}=0}^{N_s} P(M_{A}+N_A^{\downarrow},N_A^{\downarrow})$ with $M_{A}= -N_s, \ldots, N_s$.

\begin{figure*}[t!]
\includegraphics[width=1.0\linewidth]{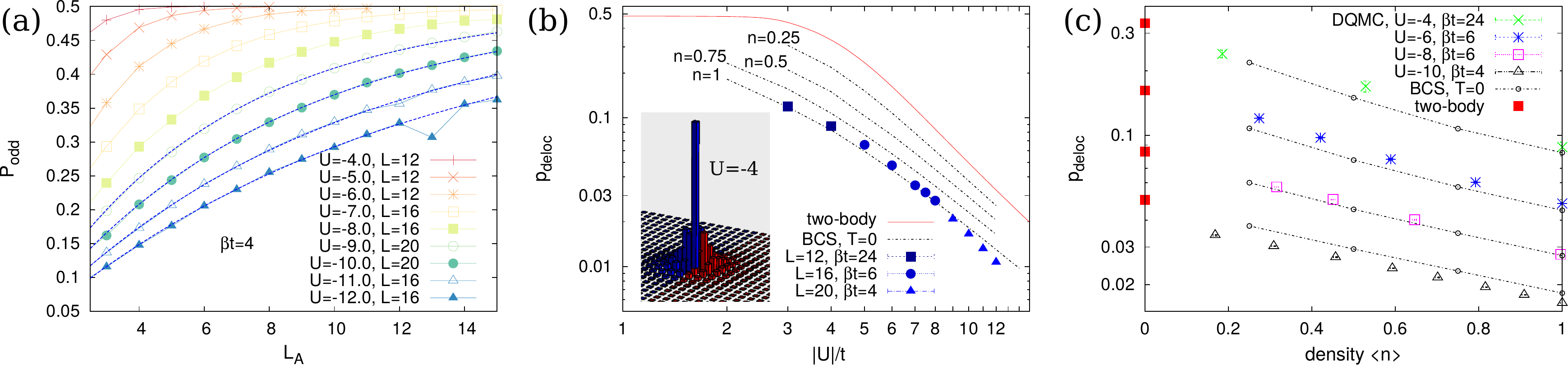}
 \caption{
  (a): Scaling of the even-odd splitting $P_{\text{odd}}$ with linear subsystem size $L_A$. Blue dotted curves 
 in (a) represent the fit function Eq.~\eqref{eq:P_odd_fit_function}.
 (b-c): Pair size $p_{\text{deloc}}$ as a function of attractive interaction $-|U|/t$ 
 and filling $\langle n \rangle$ in comparison with the predictions by mean-field theory in the BEC-BCS crossover.  
 The inset in (b) shows the two-body bound state in vacuum at $U=-4$ with red bars indicating the weight of the wave function 
 on lattice sites on the other side of the boundary. The solid red line in (b) and the red squares in (c) 
 show $p_{\text{deloc}}$ for a two-body bound state in vacuum. In the limit $\langle n \rangle \rightarrow 0$, the pair size
 as computed with DQMC approaches $p_{\text{deloc}}$ for the two-body bound state in vacuum at the same interaction $U$.}
 \label{fig:pair_size}
\end{figure*}

In the following, we demonstrate that we can efficiently determine the generating function $ \chi$ within
DQMC.
The standard procedure of DQMC \cite{Assaad2008} discretizes the inverse temperature $\beta$ 
and then decouples the interactions with a Hubbard-Stratonovich (HS)
transformation at the expense of introducing an auxiliary field at every site and time slice. Then, the 
partition function can be written as a sum over free fermion systems coupled to an (imaginary) time-dependent Ising field. 
The expectation value of an observable $\hat{O}$ is
\begin{equation}
 \langle \hat{O} \rangle = \text{Tr}\left(\hat{O} e^{-\beta \hat{H}} \right) / \text{Tr}\left( e^{-\beta \hat{H}} \right) 
  = \frac{1}{Z}\sum_{\{\vecbf{s} \}} w_{\{\vecbf{s}\}} O_{\{\vecbf{s}\}},
  \label{eq:DQMC_observable}
\end{equation}
where $Z = \sum_{\{ \vecbf{s} \}} w_{\{\vecbf{s} \}}$ is the partition sum and 
$w_{\{\vecbf{s}\}}$ is the weight of one auxiliary field configuration $\{\bf s \}$. The latter takes the form \cite{Assaad2008}
\begin{equation}
 \quad w_{\{\vecbf{s}\}} = \det \left( \mathcal{G}^{\uparrow}_{\{\vecbf{s}\}} \right)^{-1}  \left( \mathcal{G}^{\downarrow}_{\{\vecbf{s}\}} \right)^{-1},
 \label{eq:DQMC_weight}
\end{equation}
where $\mathcal{G}^{\sigma}_{\{\vecbf{s}\}} = \langle \hat{c}_{i, \sigma} \hat{c}_{j, \sigma}^{\dagger} \rangle_{\{\vecbf{s}\}}$  denotes
the single-particle Green's function for spin species $\sigma$. Such a decomposition naturally carries over to more complex quantities such
as the generating function \cite{Grover2013}
\begin{equation}
 \chi(\phi^{\uparrow},\phi^{\downarrow}) =\sum_{\{\vecbf{s} \}} w_{\{\vecbf{s}\}} \chi^{\uparrow}_{\{\vecbf{s}\}} \left(\phi^\uparrow\right)  \chi^{\downarrow}_{\{\vecbf{s}\}} \left(\phi^{\downarrow}\right) 
\end{equation}
with $\chi^{\sigma}_{\{\vecbf{s}\}} (\phi^{\sigma}) =  \text{Tr}\left( \rho^{\sigma}_{\{\vecbf{s}\}, A} \, e^{i\phi^{\sigma} \hat{N}^{\sigma}_A} \right)$
and  $\rho^{\sigma}_{\{\vecbf{s}\}, A}$ the reduced density matrix for the free fermions with auxiliary field configuration $\{\bf s \}$ and spin $\sigma$.
It is important to stress that within a Hubbard-Stratovonich configuration the reduced density matrix   factorices into a spin up and spin down part.
Each can be written  in the form of a ``Boltzmann weight'' as \cite{Chung2001, Cheong2004, Peschel2003} 
\begin{equation}
 \rho_{\{\vecbf{s}\}, A}^{\sigma} = \mathcal{K}^{\sigma}_{\{\vecbf{s}\},A} \, e^{-H_A^{\sigma}}
 \label{eq:rho_A_free_fermions}
\end{equation}
with the \emph{entanglement Hamiltonian} for spin species $\sigma$
\begin{equation}
 H_{A}^{\sigma} = - \sum_{i,j \in A} \hat{c}_{i,\sigma}^{\dagger}\log\left(\left[G^{\sigma}_{\{\vecbf{s}\},A}\right]^{-1} -\mathbb{1}\right)^{i j} \, \hat{c}_{j,\sigma}
 \label{eq:Hent}
\end{equation}
and normalization $\mathcal{K}^{\sigma}_{\{\vecbf{s}\},A}  = \det\left( \mathbb{1} - G^{\sigma}_{\{\vecbf{s}\}, A}\right)$.
Here, $\left[G^{\sigma}_{\{\vecbf{s}\},A}\right]^{ij} = \langle \hat{c}_{i,\sigma}^{\dagger} \hat{c}_{j,\sigma} \rangle_{\{\vecbf{s}\}; i,j \in A}$ is the one-body density 
matrix (OBDM) for a given HS field 
configuration $\{\vecbf{s}\}$ 
and with sites $i$ and $j$ restricted to subsystem $A$ \cite{Chung2001}. 
Therefore, the generating function for a fixed auxiliary field configuration  $\{\bf s \}$ reduces to
\begin{align}\label{eq:chi_A_Nparticles}
 \chi^{\sigma}_{\{\vecbf{s}\}}(\phi) = \prod_{\alpha=1}^{N_s} (1- \lambda^{\sigma}_{\alpha}) \prod_{\alpha=1}^{N_s} \left( 1 + e^{-\epsilon^{\sigma}_{\alpha} + i \phi} \right),
\end{align}
where $\{ \epsilon^{\sigma}_{\alpha} \}$ are the eigenvalues of the entanglement Hamiltonian $H^{\sigma}_A$
and $\{\lambda^{\sigma}_{\alpha}\}$ those of the OBDM. In deriving the above expression, we have used that 
the particle number operator $\hat{N}^{\sigma}_A = \sum_{i\in A} \hat{c}_{i,\sigma}^{\dagger} \hat{c}_{i,\sigma}$ is also
a quadratic operator, 
and commutes with the entanglement Hamiltonian $H^{\sigma}_A$, i.e., 
they have a common eigenbasis. Within this basis the grand canonical trace can be performed analytically. 

Finally,
the eigenvalues $\{ \epsilon^{\sigma}_{\alpha} \}$ and $\{\lambda^{\sigma}_{\alpha}\}$ are related
through $\epsilon^{\sigma}_{\alpha} = \log(\frac{1}{\lambda^{\sigma}_{\alpha}}-1)$ \cite{Peschel2003}, and we obtain the important result
\begin{equation}
 \chi^{\sigma}_{\{\vecbf{s}\}}(\phi) 
	    = \prod_{\alpha=1}^{N_s}(1 + (e^{i\phi} - 1)\lambda^{\sigma}_{\alpha}),
\end{equation}
which allows for the efficient determination of the generating function $\chi$ by quantum Monte Carlo sampling of the auxiliary field configurations.
If the transformations \eqref{eq:FT_chigen} and \eqref{eq:PN_marginalization}  
are performed in every Monte-Carlo measurement step, error bars for $P_N(N_A)$ and $P_M(M_A)$ 
can be obtained in the standard way. 
Note that the equal-time OBDM $G^{\sigma}_{\{\vecbf{s}\},A}(\tau)$
depends explicitly on imaginary time $\tau$, which is suppressed in our notation. 
Due to the cyclic property of the trace in Eq.~\eqref{eq:DQMC_observable} 
all imaginary time slices are equivalent and it is possible to average over them to acquire additional statistics.

In the following, we demonstrate our approach for the determination of the 
full counting statistics in the two-dimensional Fermi-Hubbard model \cite{Micnas1990} on a square lattice. 
The Hamiltonian is given by 
\begin{equation} 
 \label{eq:Hubbard_Hamiltonian}
 \mathcal{H} = - t \sum_{\langle \vecbf{r}, \vecbf{r}^{\prime} \rangle, \sigma} \left(\hat{c}_{\vecbf{r},\sigma}^{\dagger} \hat{c}_{\vecbf{r}^{\prime},\sigma} + \mathrm{h.c.} \right) +U \sum_{\vecbf{r}} \hat{n}_{\vecbf{r},\uparrow} \hat{n}_{\vecbf{r},\downarrow}.
\end{equation}
The fermionic operators $\hat{c}_{\vecbf{r},\sigma}$ obey canonical commutation relations
$\{\hat{c}^{\dagger}_{\vecbf{r},\sigma}, \hat{c}_{\vecbf{r}^{\prime},\sigma^{\prime}} \} = \delta_{\vecbf{r}, \vecbf{r}^{\prime}} \delta_{\sigma, \sigma^{\prime}}$, and  
$\hat{n}_{\vecbf{r},\sigma} = \hat{c}^{\dagger}_{\vecbf{r},\sigma} \hat{c}_{\vecbf{r},\sigma}$ are the density operators while $t$ denotes the nearest 
neighbour hopping  and $U$ the on-site interation. The model can be efficiently simulated with DQMC for the attractive case $U<0$
 at arbitrary fillings, and
for repulsive interactions at half-filling. Especially, at half filling and on bipartite lattices, a spin-down particle-hole transformation 
$\hat{c}_{\vecbf{r},{\downarrow}} \rightarrow (-1)^{r_x + r_y} \hat{c}^{\dagger}_{\vecbf{r},{\downarrow}}$
maps the attractive Hubbard model into the repulsive one.

We start with the presentation of the results for the {\it attractive Hubbard} model. The full counting statistics $P_{N}(N_{A})$ for the particle number $N_{A}$ on
the sublattice A is derived within  the setup illustrated in Fig.~\ref{fig:BEC_BCS_crossover}(a). The finite-temperature 
phase diagram \cite{Paiva2004, Paiva2010} of \eqref{eq:Hubbard_Hamiltonian} features, away from half filling, a Berezinskii-Kosterlitz-Thouless (BKT) 
transition at temperature $T_{\text{BKT}}$  to a quasi long-range $s$-wave  superconducting state. At half-filling, there is a degeneracy of s-wave superconductivity and charge density wave order, and the $SU(2)$ symmetry of the order parameter suppresses the transition temperature to zero according to the Mermin-Wagner theorm. 
In both cases, there is a second clearly separated temperature scale $T^{\star} \sim |U|$
which marks the onset of pair formation without long-range phase coherence.  
In Fig.~\ref{fig:BEC_BCS_crossover}(b-d), we present the behavior of the full counting statistics in the cross-over from the  BCS-type superfluid of large overlapping 
Cooper pairs for weak interactions to a BEC  of hardcore bosonic on-site pairs \cite{Eagles1969, Leggett1980, Nozieres1985} for strong interactions. The temperature
is chosen well below the characteristic temperature $T^{\star}$ for pair formation. 

We observe a strong even-odd effect for increasing interactions. We will argue in the following that this phenomenon can be well understood through
the size of the pairing wave function: in the extreme limit of very strong interactions, all fermions are paired up with a pair wave function localized  on a single lattice site. 
Then, we expect a  vanishing probability to find an odd number of fermions on subsystem A. An odd number of fermions can only appear due to unpaired fermions or a pairing function 
spreading over several lattice sites. The latter provides only a contribution for pairs along the boundary of subsystem A. In order to quantify this effect,
we introduce  $P_{\text{odd}} = \sum_{N_A \text{odd}} P(N_A)$
as a measure of the even-odd splitting. We expect a scaling behavior
\begin{equation}
 P_{\text{odd}}(L_A) = \frac{1}{2} - \frac{1}{2}(1 - 2 p_{\text{deloc}})^{4 L_A \cdot \langle n \rangle}
 \label{eq:P_odd_fit_function}
\end{equation}
with increasing subsystem size $L_A$ and atomic density $\langle n \rangle$; $4 L_A \cdot \langle n \rangle$ estimates
the number of pairs along the boundary. This ansatz is motivated by the idea of independent pairs randomly distributed in
the sample. Then, for a pair centered  next  to the boundary, $p_\text{deloc}$ denotes the weight of the pair wave function 
to be on the other side of the boundary. 
Therefore, we can interpret  $p_\text{deloc}$ as a measure for 
the size of the wave function.  Indeed, we find a perfect fitting of $P_{\text{odd}}$ to \eqref{eq:P_odd_fit_function} for a large parameter regime, 
see dashed lines in Fig.~\ref{fig:pair_size}(a), which allows us to extract the value  $p_{\text{deloc}}$ for different interactions and densities. 
The resulting $p_\text{deloc}$ for the Hubbard model are shown in Fig.~\ref{fig:pair_size}(b) 
at density $\langle n \rangle = 1$ and as a function of $|U|$; note 
that sufficiently low temperatures are required to suppress thermal pair breaking.
In 2D the binding energy of a bound state is exponentially small in the attractive interaction \cite{LandauLifshitz,Randeria1990}.
Therefore for $U/t=-3$, $L=12$, a temperature as low as $\beta t=24$ was necessary to achieve a fit to Eq.~\eqref{eq:P_odd_fit_function}, 
whereas for large $|U|$, e.g. $U/t = -9$, ``low enough'' temperature is $\beta t=4$. In Fig.~\ref{fig:pair_size}(c), the behavior of  $p_{\text{deloc}} $
is shown for decreasing densities.  The minimum of the pair size at half filling, $\langle n \rangle = 1$,
is a lattice effect (see \cite{Randeria1990} for the 2d continuum model), due to
the logarithmically diverging van Hove singularity in the tight-binding density of states on the square lattice.
As Fig.~\ref{fig:pair_size}(c) demonstrates, BCS theory at T=0 gives a reasonably accurate estimate of the pair size
even for intermediate values of $U$ and densities $\langle n \rangle$.

From this quantitative analysis of the even-odd splitting we conclude that the FCS can serve as a powerful tool to observe the formation
of pairing in a superconducting state and provides useful information about the size of the pairing wave function. We expect that this analysis
carries over to systems where DQMC simulations are hard, such as the repulsive Hubbard model away from half-filling, 
and can serve as a powerful experimental detection tool.

\begin{figure}[t!]
\includegraphics[width=1.0\linewidth]{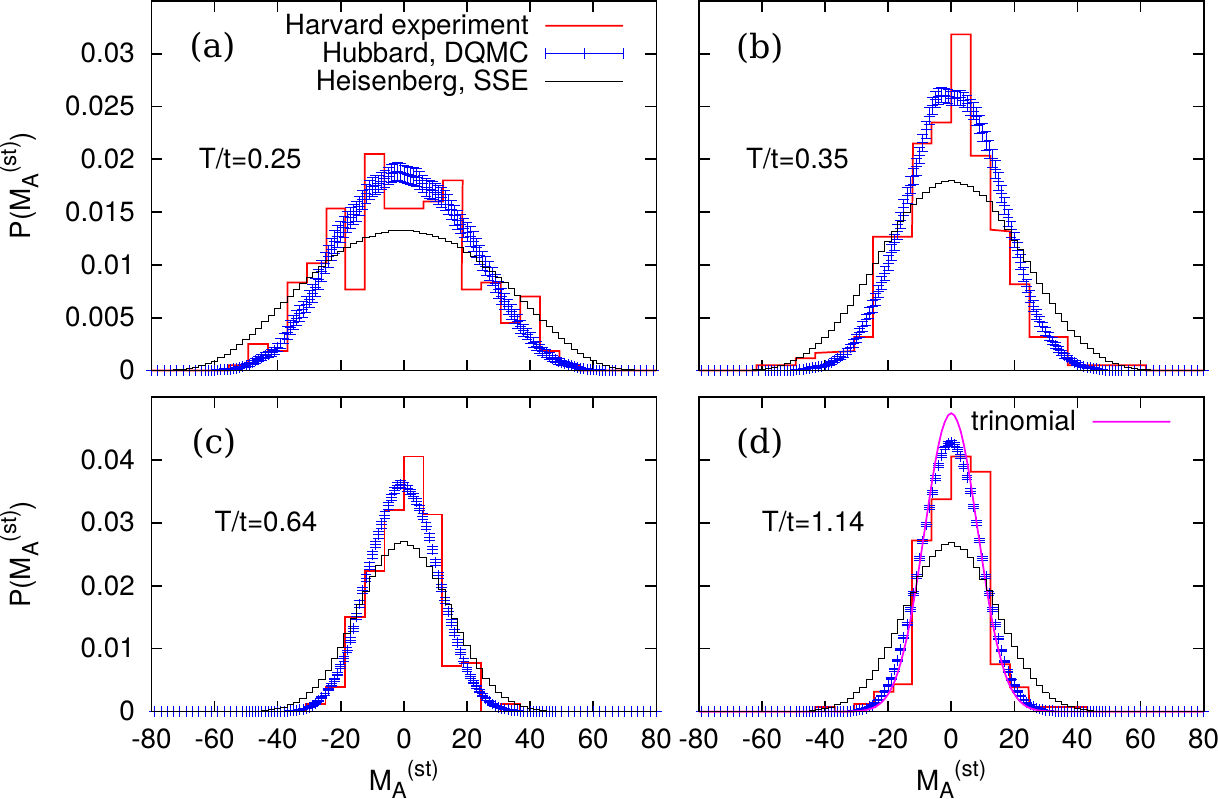}
 \caption{FCS of the staggered magnetization on a disc-shaped subsystem of $N_s=80$
 sites for repulsive Hubbard interaction $U/t = 7.2$ at half-filling. Red histograms are reproduced
 from Ref.~\cite{Mazurenko2017}, blue error bars are our DQMC simulations. Binned FCS 
 of the Heisenberg model at an equivalent temperature are shown in black. The magenta line in (d)
 is a simple model of independent sites which is parametrized by the doublon density 
 $p_d \equiv \frac{1}{N} \sum_i \langle \hat{n}_{i \uparrow} \hat{n}_{i \downarrow} \rangle = 0.066(1)$ at $T=1.14 t$, see supplement material.}
 \label{fig:Harvard_experiment}
\end{figure}

Finally, we study the FCS for the staggered magnetization in the {\it repulsive Hubbard} model at half-filling.  Recently, this model was extensively studied experimentally
using cold atoms in optical lattices \cite{Mazurenko2017} at interactions $U/t=7.2$ with a circular central region of homogeneous density 
involving approximately 80 sites surrounded by a dilute 
particle bath. Given the single-site  resolution in the experiments \cite{Parsons2016}, histograms of the staggered magnetization  
$M^{st} = \sum_i (-1)^{\vecbf{i}}(n_{i,\uparrow} - n_{i,\downarrow})$  inside the circular region were accumulated over more than 
250 experimental realizations \cite{Mazurenko2017}. Here, we show the original data points of Ref.~\cite{Mazurenko2017} as 
the red histograms in Fig.~\ref{fig:Harvard_experiment}. 
As the generating term for the staggered magnetization no longer commutes with
the entanglement Hamiltonian in Eq.~\eqref{eq:Hent} our method to determine the FCS has to be modified: it is required to explicitly determine the entanglement Hamiltonian and subsequently 
diagonalize the Hamiltonian with the staggered term of the generating function added; the details of this procedure are discussed in 
the supplement material. For optimal comparison with the experiment, we determine  the FCS of the staggered magnetization  
inside the same circular geometry as in the experiment, which is embedded in a large system of size $L \times L$ 
with $L=20$ and periodic boundary conditions. As shown in Fig.~\ref{fig:Harvard_experiment}, we find excellent agreement
without any adjustable fit parameter. Furthermore, we find strong deviation from the predictions of the staggered magnetization
from the  $S=1/2$ antiferromagnetic Heisenberg model. The latter takes the form 
$\mathcal{H}_{SO} = J \sum_{\langle i,j \rangle} \hat{\vecbf{S}}_i \cdot \hat{\vecbf{S}}_j$ with super-exchange coupling $J = 4 t^2/U$,
and follows as limiting theory of the Hubbard model for large interactions $U \gg t$; the corresponding temperature is given by $T/J\,$=$\,(T/t) U/(4t)$.
The Heisenberg model is simulated at the corresponding temperatures with the same circular geometry  of $N_s = 80$ 
sites which is embedded in a larger system of linear dimension $L=32$. For all temperatures, the half 
width of the distribution of the staggered magnetization  for the Hubbard model
is systematically smaller than that of the Heisenberg model.
We expect charge fluctuations, which already play an important role at the given interaction strength $U/t=7.2$, to
be the cause of this difference: it is well established, that the next order correction to the Heisenberg model in
$t/U$ leads to second and third neighbour exchange as well as four-spin ring exchange interactions of order $t^4/U^3$ 
\cite{MacDonald1988, Takahashi1977}. Furthermore, the operator measuring the  staggered magnetiztion is reduced by 
a renormalization factor $ \left(1 - 8\frac{t^2}{U^2}\right)$ due to doublon-hole pairs  \cite{Delannoy2005}. 
We therefore
conclude that at the intermediate interaction strength $U/t=7.2$ the differences between the Hubbard model and the Heisenberg model
already play a significant role. The perfect agreement of the experimental data with our DQMC analysis demonstrates that the
experiments are indeed capable of probing these corrections to the Heisenberg model.

However, there is an important comment of caution in order for the comparison with our theoretical predictions 
and the experimental results: As a result of the parity projection, the measurement technique of \cite{Parsons2016, Mazurenko2017} 
counts doublons and holes  sitting on the same sublattice (i.e. on diagonally opposite corners of a plaquette)
incorrectly to the staggered magnetization.  On the other hand, doublon-hole pairs on neighbouring sites, which are the 
dominant charge fluctuations, are counted correctly to the staggered magnetization as a consequence of the staggering
factor. Therefore, the experiments are currently exactly performing the measurement
to observe the leading deviations between Heisenberg and Hubbard model. However, the observation of further corrections for
lower interactions will require the precise experimental detection of the staggered magnetization.

In conclusion, we present a method to compute full quantum mechanical probability distributions of quadratic operators 
in interacting fermion systems which can be simulated with DQMC.  We find that for the attractive Hubbard model
the dependence  of an even-odd splitting in the particle number distribution function
allows one to infer the size of a preformed pair or Cooper pair from \emph{in situ} images.  Furthermore,
we apply our method to the repulsive Hubbard model, which has recently been studied experimentally 
\cite{Mazurenko2017}. The excellent agreement with our analysis demonstrates that the experiments 
are capable of observing corrections to the Heisenberg model for the intermediate interaction strengths  $U/t=7.2$.
Our method is also suitable for the study of the high-temperature phase of the repulsive Hubbard model
away from half filling or the investigation of the universality of distribution functions \cite{Lovas2017} at a quantum critical point \cite{Meng2010, Sorella2012}.

\subsection*{Acknowledgments}
We acknowledge M. Greiner for providing the raw experimental data for Figure \ref{fig:Harvard_experiment}.
S.H. thanks T. Roscilde and Feiming Hu for fruitful discussions
as well as S. Wessel for providing benchmark results in one dimension
and D. Huerga for comments on the manuscript. 
Computations were performed
on JURECA, J\"ulich Supercomputing Center. Support by the German Science Foundation (DFG) through SFB TRR21 is acknowledged.

\pagebreak

\renewcommand{\thefigure}{A\arabic{figure}}
\renewcommand{\bibnumfmt}[1]{[A#1]}
\renewcommand{\citenumfont}[1]{A#1}

\makeatletter 
\def\tagform@#1{\maketag@@@{(A.\ignorespaces#1\unskip\@@italiccorr)}}
\makeatother

\makeatletter
\makeatletter 
\makeatother

\onecolumngrid


\section{Appendix}

\subsection{FCS of the staggered particle number} 
To compute the full counting statistics (FCS) of the staggered particle number 
$N_A^{st} = \sum_{i \in A} (-1)^{\vecbf{i}} (\hat{n}_{i, \uparrow} + \hat{n}_{i,\downarrow})$
(which corresponds to the staggered magnetization  plus a constant, 
$N_A^{(st, U<0)} \rightarrow M_A^{(st, U>0)} = \sum_{i \in A} (-1)^{\vecbf{i}} (\hat{n}_{i, \uparrow} - \hat{n}_{i,\downarrow}) + \sum_{i \in A} (-1)^{\vecbf{i}} $,
in
the half-filled repulsive Hubbard model)
a modification of the method described in the main text is necessary. The generating function of $P(N_A^{st})$ in one Hubbard-Stratonovich sample is
\begin{equation}
\label{eq:chi_stN1}
 \chi_{\{\vecbf{s}\}}^{st}(\phi)  = \text{Tr}_{\uparrow} \left( e^{-H_A^{\uparrow}}  e^{i \phi \sum_{\i \in A} (-1)^{\vecbf{i}} \hat{n}_{i, \uparrow} } \right) \\
      \text{Tr}_{\downarrow} \left( e^{-H_A^{\downarrow}} e^{i \phi\sum_{i \in A}  (-1)^{\vecbf{i}} \hat{n}_{i, \downarrow}  } \right).
\end{equation}
Unlike the particle number 
$\hat{N}_A^{\sigma} = \sum_{i \in A} \hat{c}_{i, \sigma}^{\dagger} \hat{c}_{i, \sigma}$,
the operator $i \phi \sum_{i \in A} (-1)^{\vecbf{i}} \hat{c}_{i, \sigma}^{\dagger} \hat{c}_{i, \sigma}$
does not commute with the single-particle entanglement Hamiltonian $H_A^{\sigma}$ due to the staggering factor $(-1)^{\vecbf{i}}$.
Therefore, there is no common eigenbasis in which one can simply add the eigenvalues of the two operators as was done
in the last step leading to Eq. (8) of the main text. Instead, it is necessary to compute $H_A^{\sigma}$ explicitly 
in the site basis, add the non-commuting operators
\begin{equation}
 \left[ \tilde{H}_A^{\sigma} \right]_{i, j} = \left[ H_A^{\sigma} \right]_{i, j} + i \phi (-1)^{\vecbf{i}} \delta_{i, j}
 \label{eq:modified_Hent_stN}
\end{equation}
and diagonalize the resulting modified entanglement Hamiltonian $\tilde{H}_A^{\sigma}$,
which gives the eigenvalues $\tilde{\epsilon}^{\sigma}(\phi)$. 
The peculiarity of DQMC that the equal-time Green's function $G^{\sigma}_{\{\vecbf{s}\},A}$ is non-Hermitian 
and does not necessarily have a spectral decomposition 
leads to the complication that the entanglement Hamiltonian $H_A^{\sigma}$ in Eq. (7) of the main text,
\begin{equation}
 H_{A}^{\sigma} = - \sum_{i,j \in A} \hat{c}_{i,\sigma}^{\dagger}\log\left(\left[G^{\sigma}_{\{\vecbf{s}\},A}\right]^{-1} -\mathbb{1}\right)^{ij} \, \hat{c}_{j,\sigma},
 \label{eqSM:Hent}
\end{equation}
has to be computed explicitly through 
the power series of the matrix logarithm \cite{Al-Mohy2012}. This needs to be done 
only once per Hubbard-Stratonovich sample as $H_A^{\sigma}$ can be reused in \eqref{eq:modified_Hent_stN} for different values of $\phi$.

The grand-canonical trace \eqref{eq:chi_stN1} over fermionic degrees of freedom results in a single-particle determinant which 
is expressed in terms of the eigenvalues $\tilde{\epsilon}^{\sigma}(\phi)$ as
\begin{equation}
 \chi_{\{\vecbf{s}\}}^{st}(\phi) = \prod_{\sigma= \uparrow, \downarrow} \left[ \prod_{\alpha=1}^{N_{s}} (1-\lambda_{\alpha}^{\sigma})\left(1 + e^{-\tilde{\epsilon}^{\sigma}_{\alpha}(\phi)} \right)  \right].
 \label{eq:chi_stN2}
\end{equation}
The computation of the determinant \eqref{eq:chi_stN2} may be severely affected by numerical inaccuracies. 
For low temperatures the occupation numbers $\lambda^{\sigma}_{\alpha}$ in the free fermion system tend to zero or one and,
consequently, the OBDM $G^{\sigma}_{\{\vecbf{s}\},A}$ is a nearly singular matrix. The evaluation of \eqref{eqSM:Hent} in finite-precision arithmetic
is beset with numerical instabilities and, even if a high-quality implementation of the matrix logarithm \cite{Al-Mohy2012} is used,
not all eigenvalues $\epsilon^{\sigma}_{\alpha}$ of the entanglement Hamiltonian $H_A^{\sigma}$ can be obtained with sufficient accuracy across the entire spectrum.
Consider small (negative) ``entanglement energy'' $\epsilon^{\sigma}_{\alpha}$ for which according to the Fermi-Dirac statistics 
$\lambda^{\sigma}_{\alpha} = \frac{1}{1 + e^{\epsilon^{\sigma}_{\alpha}}} \approx 1$.
The evaluation of $\epsilon^{\sigma}_{\alpha} = \log \left(\frac{1}{\lambda^{\sigma}_{\alpha}} -1\right) \approx \log(1 - 1)$, or equivalently 
the evaluation of \eqref{eqSM:Hent}, 
is very inaccurate for the small (negative) real part of the entanglement spectrum since $\log(x)$ varies greatly for $x \rightarrow 0^{+}$. 
In view of round-off and cancellation errors it is therefore  crucial to compute \eqref{eqSM:Hent} with the mathematically equivalent
formula 
\begin{equation*}
 \left[H_A^{\sigma}\right]_{i,j} = - \left[\log\left( G^{\sigma}_{\{ \vecbf{s}\},A} (\mathbb{1} - G^{\sigma}_{\{ \vecbf{s}\},A})^{-1} \right)\right]_{i,j}.
\end{equation*}
This expression shifts the accuracy to small (negative)
$\epsilon^{\sigma}_{\alpha}$ (large $\lambda^{\sigma}_{\alpha} \approx 1$) since $\log(x)$
is not much affected by errors in its \emph{large} argument $x = \frac{\lambda^{\sigma}_{\alpha}}{1 - \lambda^{\sigma}_{\alpha}}$. 
For the computation of \eqref{eq:chi_stN2} errors in the large (positive) part of the entanglement spectrum
$\epsilon^{\sigma}_{\alpha}$ are not problematic because large $\tilde{\epsilon}^{\sigma}_{\alpha}(\phi)$ are irrelevant in the factor
$(1 - \lambda^{\sigma}_{\alpha})\left( 1 + e^{-\tilde{\epsilon}^{\sigma}_{\alpha}(\phi)}\right)$. 
Here it is assumed that the spectrum $\tilde{\epsilon}^{\sigma}_{\alpha}(\phi)$ 
of the modified entanglement Hamiltonian \eqref{eq:modified_Hent_stN}
is qualitatively similar to that of the entanglement Hamiltonian \eqref{eqSM:Hent} in the sense that the accuracy of the relevant small
$\tilde{\epsilon}^{\sigma}_{\alpha}(\phi)$ is not affected by the inaccuracy of large $\epsilon^{\sigma}_{\alpha}$.

\subsection{FCS for a BCS mean-field state}

The Hamiltonian of the single-band Fermi-Hubbard model in momentum space reads
\begin{equation}
  \label{eq:Hubbard_momentum}
 \mathcal{H}_{U<0} = \sum_{\vecbf{k},\sigma} (\varepsilon_{\vecbf{k}} - \mu) \hat{c}_{\vecbf{k},\sigma}^{\dagger} \hat{c}_{\vecbf{k},\sigma} 
    \\ -\frac{|U|}{N_{\text{sites}}} \sum_{\vecbf{k}, \vecbf{k^{\prime}}, \vecbf{q}} 
    \hat{c}_{\vecbf{k} + \vecbf{q}, \uparrow}^{\dagger} \hat{c}_{-\vecbf{k} + \vecbf{q}, \downarrow}^{\dagger}
    \hat{c}_{\vecbf{k^{\prime}} + \vecbf{q}, \uparrow} \hat{c}_{-\vecbf{k^{\prime}} + \vecbf{q}, \downarrow},
\end{equation} 
with the single-particle band structure $\varepsilon_{\vecbf{k}} = -2t (\cos(k_x a) + \cos(k_y a))$ for the square lattice.
The standard BCS mean-field analysis starts with the BCS \emph{reduced} Hamiltonian which neglects 
the sum over $\vecbf{q}$ in \eqref{eq:Hubbard_momentum} and considers only scattering 
between pairs of zero center-of-mass momentum ($\vecbf{q} = 0$). 
Under the assumption of a non-zero $s$-wave pairing order parameter 
$\Delta = \sum_{\vecbf{k^{\prime}}} \langle \hat{c}_{\vecbf{k},\uparrow} \hat{c}_{-\vecbf{k},\downarrow} \rangle$
a mean-field decoupling is performed. The resulting quadratic Hamiltonian is then 
solved with a Bogoliubov transformation \cite{Schrieffer_book} with coefficients \begin{equation}
 v_{\vecbf{k}}^2 = 1 - u_{\vecbf{k}}^2 = \frac{1}{2}\left( 1 - \frac{\varepsilon_{\vecbf{k}}^{(\text{HF})} - \mu(n,U)}{ E_{\vecbf{k}}} \right),
 \label{eq:BCS_coeff}
\end{equation}
which are parametrized by the chemical potential $\mu$ and $\Delta$, the gap parameter.
$E_{\vecbf{k}}^2 = (\varepsilon_{\vecbf{k}}^{(\text{HF})} - \mu(n,U))^2 + \Delta^2(n,U)$ 
are the excitation energies of the fermionic Bogoliubov quasiparticles.
In order to take into account density-density interactions at the mean-field level, it is necessary to include 
the Hartree-Fock potential in the 
single-particle energies: 
$\varepsilon_{\vecbf{k}}^{(\text{HF})} = \varepsilon_{\vecbf{k}} - |U| \frac{n}{2}$
where $n = (N_{\uparrow} + N_{\downarrow}) / N_{\text{sites}}$ is the filling.
The mean-field ground state takes the standard form in terms of the BCS coefficients \eqref{eq:BCS_coeff}
\begin{equation}
 |\text{BCS}\rangle = \prod_{\vecbf{k}}\left( u_{\vecbf{k}}(\mu,\Delta) + v_{\vecbf{k}}(\mu,\Delta) \hat{c}_{\vecbf{k}, \uparrow}^{\dagger} \hat{c}_{-{\vecbf{k}}, \downarrow}^{\dagger} \right) | \text{vac} \rangle,
 \label{eq:BCS_wavefunction}
\end{equation}
where $\mu(n, U)$ and $\Delta(n,U)$ are self-consistent solutions \cite{Leggett1980_SM}
of the gap equation 
\begin{equation}
 \frac{1}{|U|} = \frac{1}{N_{\text{sites}}} \sum_{\vecbf{k} \in \text{1st BZ}} \frac{1}{2 \sqrt{ (\varepsilon_{\vec{k}} - \mu - |U| \frac{n}{2})^2 + \Delta^2}},
 \label{eq:gap_eq}
\end{equation}
and the number equation 
\begin{equation}
 n = 1 - \frac{1}{N_{\text{sites}}} \sum_{\vecbf{k} \in \text{1st BZ}} \frac{\varepsilon_{\vec{k}} - \mu - |U| \frac{n}{2}}{ \sqrt{ (\varepsilon_{\vec{k}} - \mu - |U| \frac{n}{2})^2 + \Delta^2}}.
 \label{eq:number_eq}
\end{equation}
For general filling $n$ and Hubbard interaction $U$ these self-consistent equations need to be solved numerically
to obtain $\mu$ and $\Delta$.
At half filling ($n=1$) the number equation has the solution $\mu = -|U|/2$ for any choice of $\Delta$
provided that the single-particle dispersion relation $\varepsilon_{\vecbf{k}}$ is particle-hole
symmetric around $\varepsilon_{|\vecbf{k}|=k_F} = 0$ such that the integral in \eqref{eq:number_eq} vanishes. 
Thus, the inclusion of the Hartree shift in the single-particle energies has preserved, 
at the mean-field level, the particle-hole symmetry which the Hubbard model posseses for half 
filling. In the atomic limit $t/U=0$ the solutions are 
$\mu = -|U|/2$ and $\Delta = |U| \sqrt{n(2-n)}/2$.

The BCS state is the ground state of a quadratic (mean-field) Hamiltonian and therefore Wick's theorem
can be used to factorize any correlation function into sums of products of single-particle Green's functions.
Following closely the method of Ref.~\cite{Cherng2007_SM}, the generating function for the total particle number is 
written as 
\begin{equation}
 \chi^{(N)}(\phi) = \langle e^{i \phi \hat{N}_A}\rangle = \langle \prod_{i \in A} \prod_{\sigma=\uparrow, \downarrow} e^{i \phi (1 - \hat{n}_{i, \sigma})} \rangle
                    = \langle \prod_{i \in A} \prod_{\sigma=\uparrow, \downarrow} E_{i \sigma} F_{i \sigma}(\phi) \rangle,
  \label{eq:BCS_Wicks_theorem}                  
\end{equation}
where $E_{i \sigma} = \hat{c}_{i \sigma} + \hat{c}_{i \sigma}^{\dagger}$ and $F_{i \sigma} = \hat{c}_{i \sigma} + e^{i \phi} \hat{c}_{i \sigma}^{\dagger}$.
In the second equation of \eqref{eq:BCS_Wicks_theorem} the particle-hole symmetry of the Hubbard Hamiltonian
was used to replace the total particle number by the total hole number, which simplifies the analysis \cite{Cherng2007_SM}
as the identity $e^{i \lambda (1 - \hat{c}_i^{\dagger} \hat{c}_i)} = E_i F_i(\phi)$ can then be used.
The multipoint correlation function \eqref{eq:BCS_Wicks_theorem} is contracted according to Wick's theorem,
and, taking into account the fermionic anticommutation relations, the non-zero full contractions \cite{Cherng2007_SM}
constitute a determinant:
\begin{align}
  \chi^{(N)}(\phi) &= \det_{\substack{i,j=1,N_{s} \\ \sigma, \sigma^{\prime} = \uparrow, \downarrow}} \left( \langle E_{i \sigma} F_{j \sigma^{\prime}}(\phi) \rangle\right) \\
		   \label{eq:chi_Bernoulli}
                   &= \prod_{k=1}^{2 N_s} (\mu_k + (1 - \mu_k) e^{i \phi}).
\end{align}
In the last step we have introduced the eigenvalues $\mu_k$ of the matrix of normal and anomalous Green's functions:
\begin{equation}
 M = \begin{pmatrix}
                \langle \hat{c}_{i\uparrow}^{\dagger} \hat{c}_{j \uparrow}\rangle_{i,j \in A}  & \langle \hat{c}_{i \uparrow } \hat{c}_{j \downarrow}\rangle_{i,j \in A}  \\
                \langle \hat{c}_{i \downarrow} \hat{c}_{j \uparrow }\rangle_{i,j \in A}  & \langle \hat{c}_{i \downarrow}^{\dagger} \hat{c}_{j \downarrow} \rangle_{i,j \in A}   \\
               \end{pmatrix}.
\end{equation}
The Green's functions, which are restricted to subsystem $A$, are computed for the BCS state \eqref{eq:BCS_wavefunction}.
As described in detail in \cite{Cherng2007_SM}, complex conjugate pairs of eigenvalues of $M$ can interfere in Eq.~\eqref{eq:chi_Bernoulli}
leading to a suppression of odd versus even values in the particle number distribution $P(N_A)$.
Complex eigenvalues can appear as soon as non-zero values of the anomalous Green's functions
$\langle \hat{c}_{i \uparrow } \hat{c}_{j \downarrow}\rangle_{i,j \in A} = - \langle \hat{c}_{j \downarrow} \hat{c}_{i \uparrow } \rangle_{i,j \in A}\neq 0$
destroy the hermiticity of $M$, which makes evident that BCS pairing is at the origin of the even-odd effect in 
the distibution $P(N_A)$.

With increasing $|U|$ the 
local number fluctuations of the BCS state \eqref{eq:BCS_wavefunction} acquire a large unphysical extensive (Poissonian) contribution
(see Figs. 1(c) and (e) of the main text),
which is partly due to the fact that the mean-field Hamiltonian does not conserve the total number of particles. 
Furthermore, the wave function \eqref{eq:BCS_wavefunction}, 
when applied to a lattice model, fails to account for the nearest-neighbour
repulsion \cite{Micnas1990_SM} of tightly bound pairs in the limit of strong correlations $|U| \gg t $,
where interactions between Bogoliubov quasiparticles need to be included \cite{Belkhir1992_SM}.
This is different from the equivalent model in the continuum where \eqref{eq:BCS_wavefunction}
becomes the exact ground state wavefunction both in the extreme BCS and BEC limit \cite{Eagles1969_SM,Leggett1980_SM,Nozieres1985_SM}
and reproduces the correct particle number variance and higher cumulants \cite{Belzig2007_SM}.

\subsection{Trinomial distribution}
\label{subsec:binotrino}

A simple model that is capable of describing the distribution of the staggered magnetization 
$M_A^{st}$ in the limit of high temperatures (see Fig. 3(d) of the main text) regards the staggered magnetization at each lattice site 
as an independent random variable $m_i^{st}$ which can take on the values $0,+1,-1$ with 
the probabilities
\begin{align}
 p_1 \equiv p(m_i^{st}=+1) = p(\uparrow, +) + p(\downarrow,-) \nonumber \\
 p_2 \equiv p(m_i^{st}=-1) = p(\uparrow, -) + p(\downarrow,+) \\
 p_3 \equiv p(m_i^{st}=0) = p(d) + p(h) = 1 - p_1 - p_2, \nonumber
\end{align}
with $p(\sigma, f)$ denoting the probability of the elementary event that a spin $\sigma \in (\uparrow, \downarrow)$ is 
placed on a lattice site with staggering factor $(-1)^{\vecbf{i}} \equiv f \in (+,-)$ and $d$ and $h$ signifying the placement 
of a doublon and hole, respectively. In the presence of particle-hole symmetry and equal chemical
potentials for spin up and down, $p(d) = p(h)$ and $p(\uparrow, f) = p(\downarrow, -f)$,
such that the probabilities of all three elementary events are described by the single parameter $p_d$,
which we set to the average double occupancy $p_d \leftarrow \langle d \rangle = \frac{1}{N} \sum_i \langle \hat{n}_{i \uparrow} \hat{n}_{i \downarrow} \rangle$
as computed with Monte Carlo. The total staggered magnetization on a subsystem with $N_s$ sites is given
by a sum over a trinomial distribution
\begin{equation}
 P(M^{st}_A =k-l) = \sum_{k=0}^{N_s} \sum_{l=0}^{N_s-k} \delta_{M^{st}_A, k-l} P_3(X=k, Y=l)
 \label{eq:trinomial}
\end{equation}
where 
\begin{equation}
 P_3(X=k, Y=l)   = \frac{N_s !}{k!\, l!\, \left(N_s - k - l \right)!} \, {p_1}^{k} {p_2}^{l} (1 - p_1 - p_2)^{N_s - k - l}.
\end{equation}
For the distribution function in Fig. 3(d) of the main text, the value $\langle d \rangle = 0.066(1)$ as extracted
from the Monte Carlo simulations at temperature $T/t=1.14$ was used.

\subsection{ Additional datasets: FCS of magnetization, small subsystems}

\begin{figure}[h!]
 \includegraphics[width=0.75\linewidth]{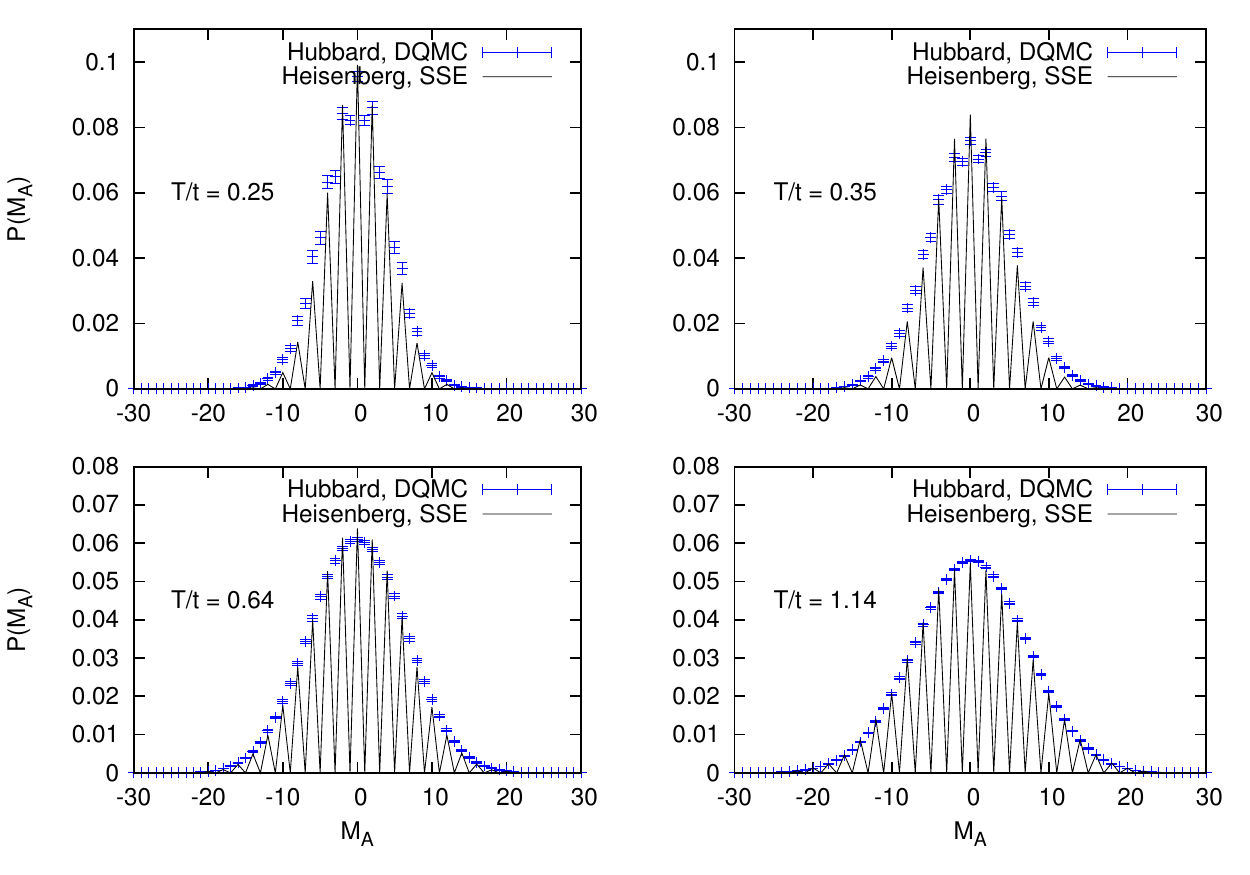}
 \caption{FCS of the magnetization $\hat{M}_A = \sum_{i \in A}(\hat{n}_{i,\uparrow} - \hat{n}_{i,\downarrow})$ 
 in the repulsive Hubbard model at half filling for a disc-shaped subsystem $A$ of $N_s=80$ sites. The parameters 
 are the same as in Fig. 3 of the main text.}
 \label{fig:80sites_disc_magnetization}
\end{figure}

The FCS of the magnetization $\hat{M}_A = \sum_{i \in A}(\hat{n}_{i,\uparrow} - \hat{n}_{i,\downarrow})$ (see Fig.~\ref{fig:80sites_disc_magnetization}) agrees much better between Hubbard and Heisenberg model than that of the staggered magnetization.
This can be traced back to the fact that the total magnetization $\hat{M}^{\text{tot}} = \sum_i (\hat{n}_{i \uparrow} - \hat{n}_{i \downarrow})$
commutes with the Hubbard Hamiltonian $\mathcal{H}$ and the magnetization on a subsystem $A$
commutes up to an operator $\hat{O}_{\partial A}$ that has support only on the boundary of $A$, $[\mathcal{H}, \hat{M}_A] = \hat{O}_{\partial A}$,
which comes from the change of magnetization due to particles hopping into and out of the subsystem. A possible renormalization factor 
of the magnetization
that derives from the canonical transformation of the Hubbard model into a spin-only Hamiltonian \cite{Delannoy2005_SM}
should then be at most a boundary effect. 
For temperatures $T/t=0.25$ and $T/t=0.35$ a tiny even-odd modulation is visible in the FCS of the magnetization.
This is the precursor of the full suppression of odd magnetizations in the FCS of the Heisenberg model.
Unlike in Fig. 3 of the main text, in Fig.~\ref{fig:80sites_disc_magnetization} 
odd magnetizations have not been binned together with the nearest even magnetization
to give a smooth histogram. Instead, the probability distribution $P(M_A)$ for the Heisenberg model 
has been divided by the bin width, i.e. by two, to aid the visual comparison
with the probability distribution for the Hubbard model which has twice as many 
possible values for $M_A$ on the abscissa.

\begin{figure}[h!]
\includegraphics[width=1.0\linewidth]{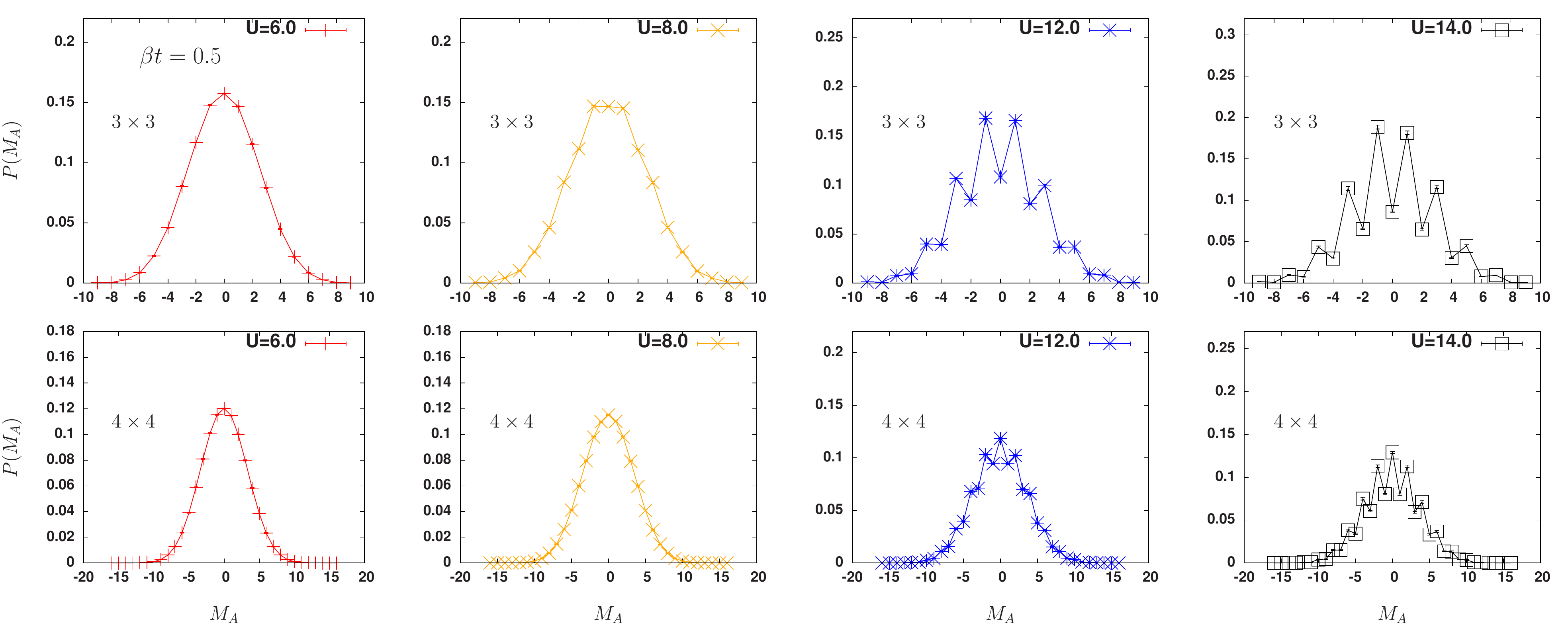}
\includegraphics[width=1.0\linewidth]{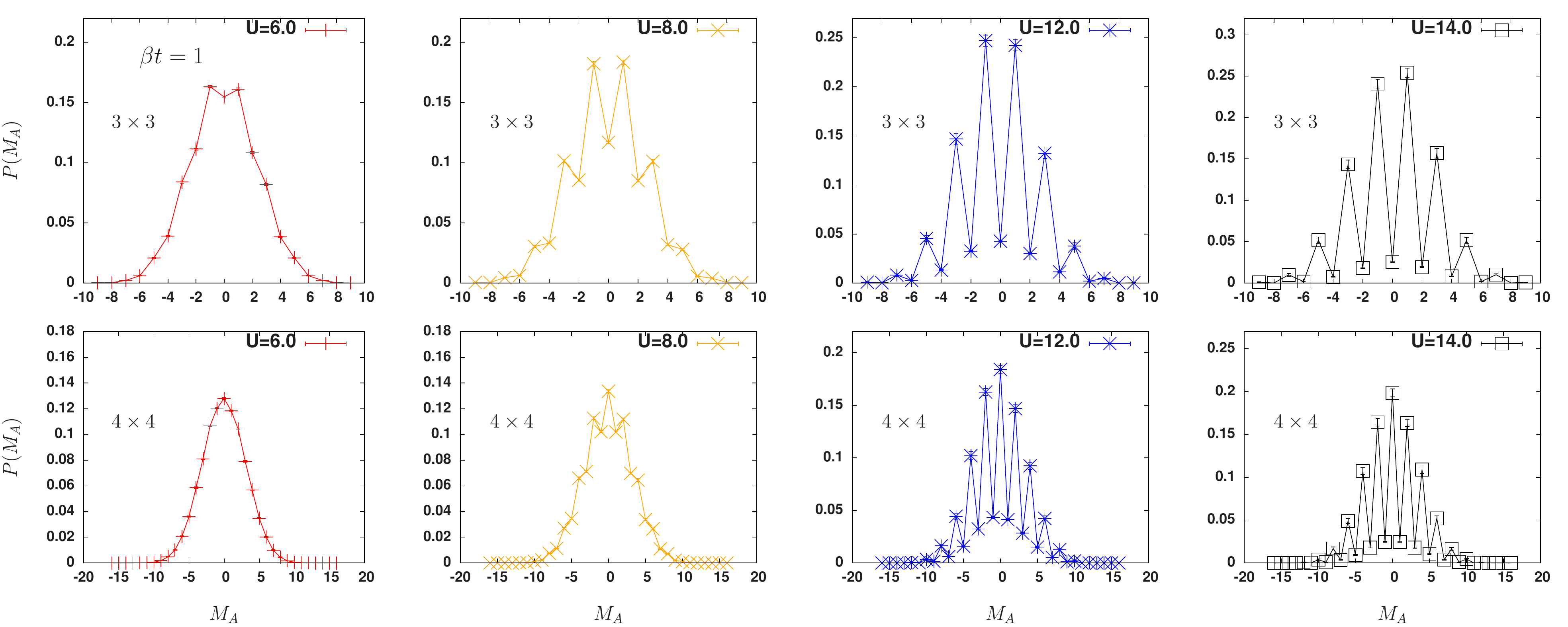}
\includegraphics[width=1.0\linewidth]{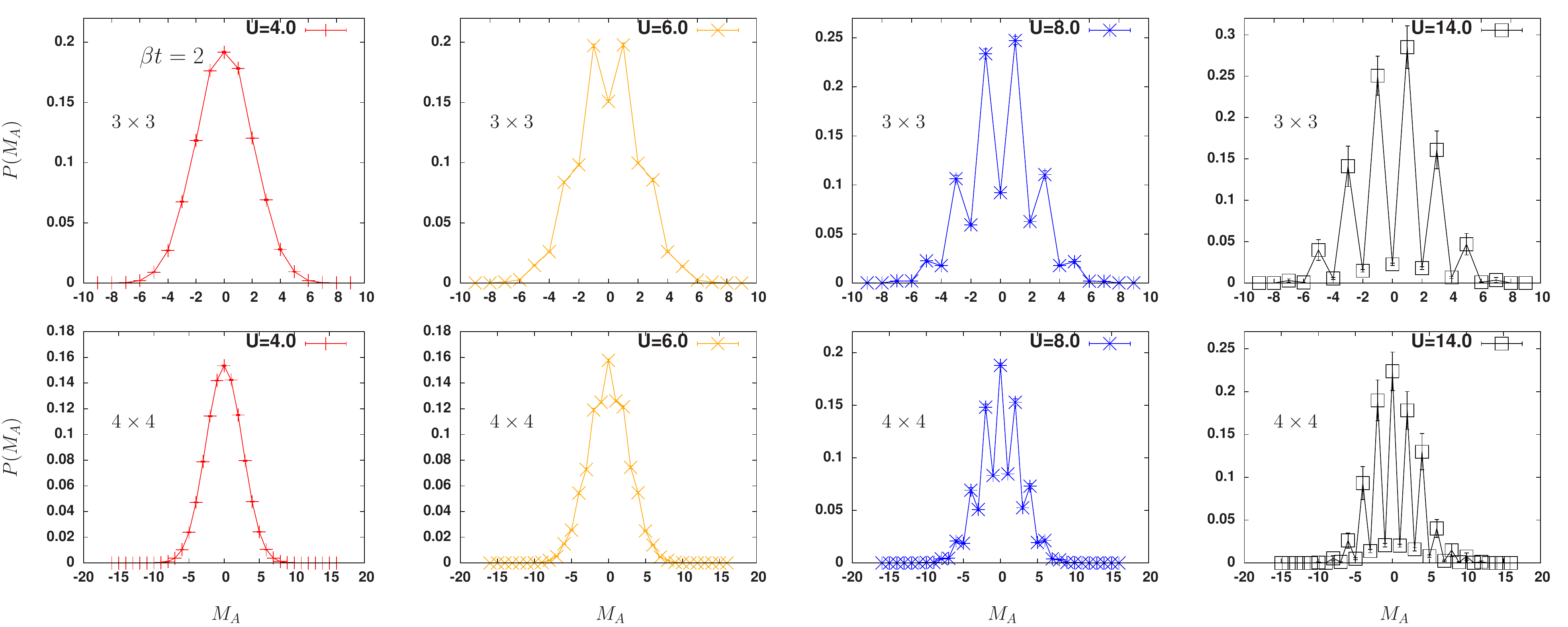}
 \caption{FCS of the magnetization $\hat{M}_A = \sum_{i \in A}(\hat{n}_{i,\uparrow} - \hat{n}_{i,\downarrow})$
 in the repulsive Hubbard model at half filling 
 for $N_s = 3 \times 3$
 and $N_s = 4 \times 4$ subsystems.
 The total system size is $16 \times 16$ lattice sites with periodic boundary conditions.
 The inverse temperature is $\beta t=0.5$ (upper two rows), $\beta t=1$ (middle two rows), 
 and $\beta t=2$ (lowest two rows).  }
 \label{fig:PNA3x3and4x4}
\end{figure}
\begin{figure}-
 \includegraphics[width=0.9\linewidth]{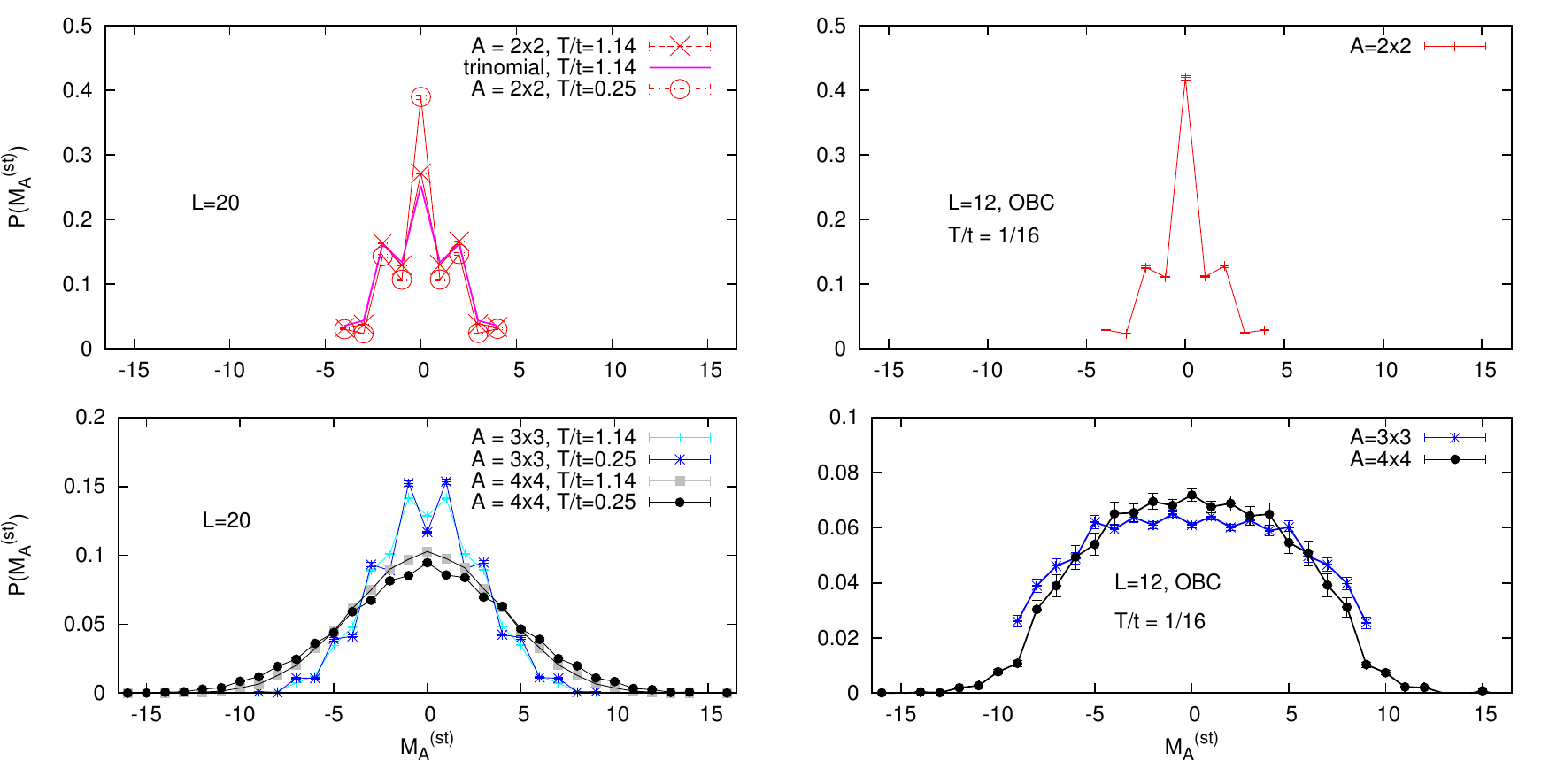}
 \caption{FCS of the staggered magnetization 
 $\hat{M}^{st}_A = \sum_{i \in A}(-1)^{\bf i}(\hat{n}_{{\bf i},\uparrow} - \hat{n}_{{\bf i},\downarrow})$ 
 in the repulsive Hubbard model at half filling on small subsystems
 of $N_s = 2 \times 2, 3 \times 3$, and $4 \times 4$ sites.
 Left column: FCS for the highest and lowest temperature of Ref.~\cite{Mazurenko2017_SM}.
 Total linear system size $L=20$ with periodic boundary conditions.
 The magenta line is calculated from a trinomial distribution according to Eq.~\eqref{eq:trinomial}.
 The only parameter entering this calculation is the double occupancy $p_d = 0.0661$ (see previous section),
 which was extracted from the Monte Carlo simulations at temperature $T/t=1.14$.
 Right column: FCS for a low temperature $T/t=1/16=0.0625$. Linear system size $L=12$
 with open boundary conditions (OBC). The probe areas $A$ are located at the center of the $L \times L$
 square with OBC.}
 \label{fig:PstM_2x2_3x3_4x4}
\end{figure}

In Fig.~\ref{fig:PNA3x3and4x4} the distribution $P(M_A)$ of the magnetization 
in the repulsive Hubbard model 
at half filling is shown for small subsystems of size $3 \times 3$ and $4 \times 4$.
The interaction strength ranges from $U=4$ to $U=14$ and the inverse temperature from $\beta t = 0.5$
to $\beta t = 2$. Already at these high temperatures an even-odd splitting is visible 
for large enough Hubbard repulsion $U$.

Fig.~\ref{fig:PstM_2x2_3x3_4x4} shows the distribution $P(M^{st}_A)$ of the staggered magnetization  
$\hat{M}^{st}_A = \sum_{i \in A}(-1)^{\vecbf{i}}(\hat{n}_{i,\uparrow} - \hat{n}_{i,\downarrow})$ 
in the half-filled repulsive Hubbard model at $U/t=7.2$ for small subsystems $A$.
In the left column of Fig.~\ref{fig:PstM_2x2_3x3_4x4}, the FCS is shown 
for the highest and lowest temperatures realized in the experiment of Ref.~\cite{Mazurenko2017_SM}.
For the highest temperature $T/t=1.14$, we find good agreement 
with a simple model based on the trinomial distribution (magenta line) which treats 
each lattice site as independent (see previous section) and is parametrized solely by the average double occupancy
$p_d = \frac{1}{N} \sum_{\vecbf{r}} \langle \hat{n}_{\vecbf{r}, \uparrow} \hat{n}_{\vecbf{r}, \downarrow} \rangle$ = 0.066(1)
as computed with Monte Carlo for the given temperature.
Interestingly, the even-odd splitting in $P(M^{st}_A)$ is smeared out much more quickly 
with increasing subsystem size than in the case of $P(M_A)$. The origin 
of the even-odd asymmetry is in both cases the formation of local 
moments; however, contrary to $M_A$ ($N_A$ for $\mathcal{H}_{U<0}$ and half filling), which can only change 
by $\Delta M_A = \pm 1$ when a particle crosses the boundary of subsystem $A$, 
$M^{st}_A$ ($N^{st}_A$ for $\mathcal{H}_{U<0}$ and half filling)
is additionally affected by hopping processes in the bulk of $A$,
which change the staggered magnetization by $\Delta M^{st}_A = \pm 2$.
At this level of analysis and at half filling, the even-odd splitting in
$P(M^{st}_A)$ provides no additional information 
beyond the parameter $p_{\text{deloc}}$ extracted earlier from the even-odd splitting
in $P(M_A)$ (or rather $P(N_A)$ for $H_{U<0}$).

\subsection{Benchmarking}

\begin{figure}[h!]
 \includegraphics[width=0.5\linewidth]{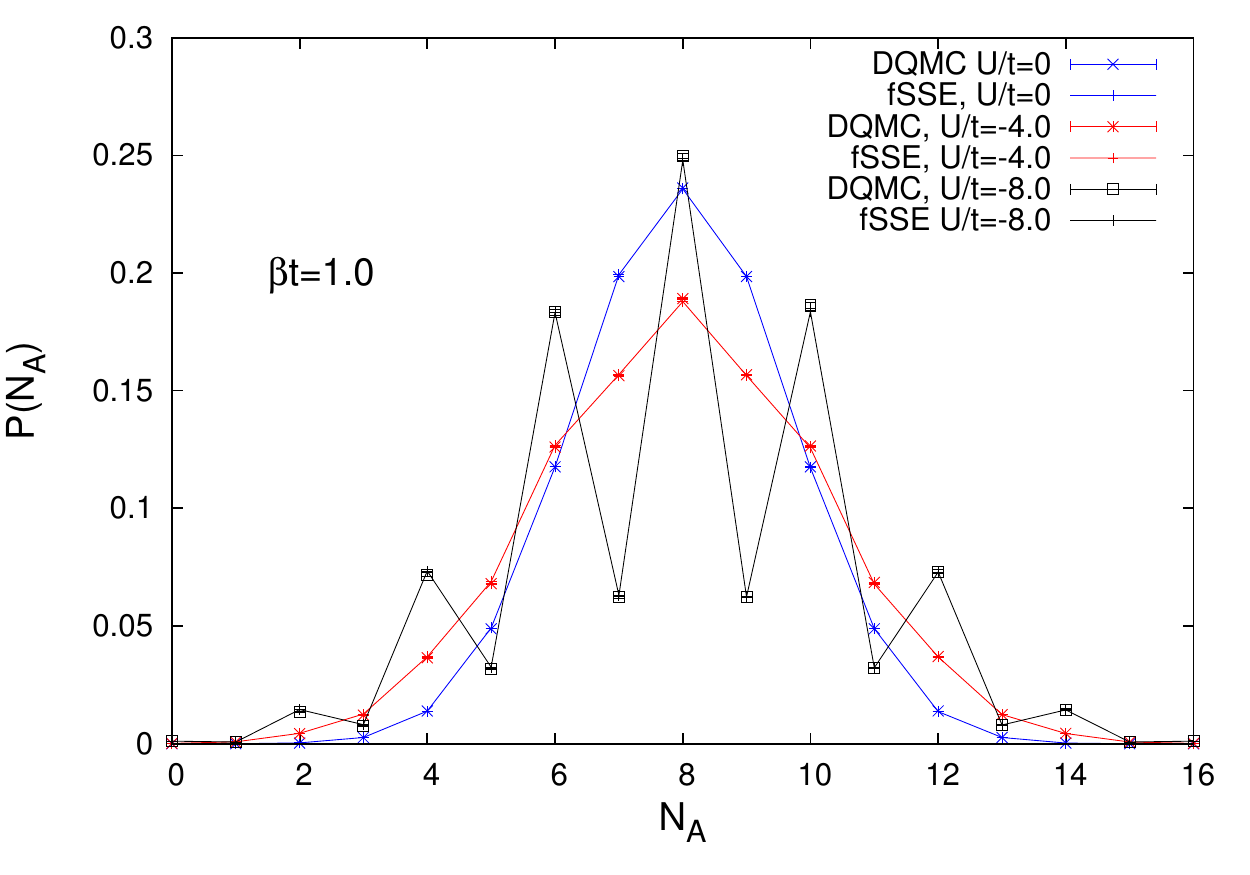}
 \caption{Comparison of our DQMC method with Stochastic Series Expansion (SSE) QMC for a Fermi-Hubbard chain (fSSE)  of $L = 16$ 
 sites with open boundary conditions at inverse temperature $\beta t = 1.0$; the subsystem size is $L_A = 8$.}
 \label{fig:benchmarking}
\end{figure}

In order to verify the correctness of our numerical implementation 
we compare our DQMC method  with Stochastic Series Expansion  \cite{Sandvik1999_SM} (SSE) QMC for a Fermi-Hubbard chain \cite{Sandvik1992_SM},
where the FCS can be obtained simply by accumulating histograms over Monte Carlo measurement steps. 
Open boundary conditions are necessary to ensure that the 
SSE QMC has no sign problem. The agreement between the two methods is excellent (see Fig.~\ref{fig:benchmarking}).

%


\begin{thebibliography}{39}%
\makeatletter
\providecommand \@ifxundefined [1]{%
 \@ifx{#1\undefined}
}%
\providecommand \@ifnum [1]{%
 \ifnum #1\expandafter \@firstoftwo
 \else \expandafter \@secondoftwo
 \fi
}%
\providecommand \@ifx [1]{%
 \ifx #1\expandafter \@firstoftwo
 \else \expandafter \@secondoftwo
 \fi
}%
\providecommand \natexlab [1]{#1}%
\providecommand \enquote  [1]{``#1''}%
\providecommand \bibnamefont  [1]{#1}%
\providecommand \bibfnamefont [1]{#1}%
\providecommand \citenamefont [1]{#1}%
\providecommand \href@noop [0]{\@secondoftwo}%
\providecommand \href [0]{\begingroup \@sanitize@url \@href}%
\providecommand \@href[1]{\@@startlink{#1}\@@href}%
\providecommand \@@href[1]{\endgroup#1\@@endlink}%
\providecommand \@sanitize@url [0]{\catcode `\\12\catcode `\$12\catcode
  `\&12\catcode `\#12\catcode `\^12\catcode `\_12\catcode `\%12\relax}%
\providecommand \@@startlink[1]{}%
\providecommand \@@endlink[0]{}%
\providecommand \url  [0]{\begingroup\@sanitize@url \@url }%
\providecommand \@url [1]{\endgroup\@href {#1}{\urlprefix }}%
\providecommand \urlprefix  [0]{URL }%
\providecommand \Eprint [0]{\href }%
\providecommand \doibase [0]{http://dx.doi.org/}%
\providecommand \selectlanguage [0]{\@gobble}%
\providecommand \bibinfo  [0]{\@secondoftwo}%
\providecommand \bibfield  [0]{\@secondoftwo}%
\providecommand \translation [1]{[#1]}%
\providecommand \BibitemOpen [0]{}%
\providecommand \bibitemStop [0]{}%
\providecommand \bibitemNoStop [0]{.\EOS\space}%
\providecommand \EOS [0]{\spacefactor3000\relax}%
\providecommand \BibitemShut  [1]{\csname bibitem#1\endcsname}%
\let\auto@bib@innerbib\@empty
\bibitem [{\citenamefont {Levitov}\ \emph {et~al.}(1996)\citenamefont
  {Levitov}, \citenamefont {Lee},\ and\ \citenamefont {Lesovik}}]{Levitov1996}%
  \BibitemOpen
  \bibfield  {author} {\bibinfo {author} {\bibfnamefont {L.~S.}\ \bibnamefont
  {Levitov}}, \bibinfo {author} {\bibfnamefont {H.}~\bibnamefont {Lee}}, \ and\
  \bibinfo {author} {\bibfnamefont {G.~B.}\ \bibnamefont {Lesovik}},\
  }\href@noop {} {\bibfield  {journal} {\bibinfo  {journal} {Journal of
  Mathematical Physics}\ }\textbf {\bibinfo {volume} {37}},\ \bibinfo {pages}
  {4845} (\bibinfo {year} {1996})}\BibitemShut {NoStop}%
\bibitem [{Qua()}]{QuantumNoise}%
  \BibitemOpen
  \href@noop {} {}\bibinfo {note} {\emph{Quantum Noise in Mesoscopic Physics},
  edited by Yu. V. Nazarov (Kluwer, Dordrecht 2003)}\BibitemShut {NoStop}%
\bibitem [{\citenamefont {{de-Picciotto}}\ \emph {et~al.}(1998)\citenamefont
  {{de-Picciotto}}, \citenamefont {{Reznikov}}, \citenamefont {{Heiblum}},
  \citenamefont {{Umansky}}, \citenamefont {{Bunin}},\ and\ \citenamefont
  {{Mahalu}}}]{de-Picciotto1998}%
  \BibitemOpen
  \bibfield  {author} {\bibinfo {author} {\bibfnamefont {R.}~\bibnamefont
  {{de-Picciotto}}}, \bibinfo {author} {\bibfnamefont {M.}~\bibnamefont
  {{Reznikov}}}, \bibinfo {author} {\bibfnamefont {M.}~\bibnamefont
  {{Heiblum}}}, \bibinfo {author} {\bibfnamefont {V.}~\bibnamefont
  {{Umansky}}}, \bibinfo {author} {\bibfnamefont {G.}~\bibnamefont {{Bunin}}},
  \ and\ \bibinfo {author} {\bibfnamefont {D.}~\bibnamefont {{Mahalu}}},\
  }\href@noop {} {\bibfield  {journal} {\bibinfo  {journal} {Physica B
  Condensed Matter}\ }\textbf {\bibinfo {volume} {249}},\ \bibinfo {pages}
  {395} (\bibinfo {year} {1998})}\BibitemShut {NoStop}%
\bibitem [{\citenamefont {Saminadayar}\ \emph {et~al.}(1997)\citenamefont
  {Saminadayar}, \citenamefont {Glattli}, \citenamefont {Jin},\ and\
  \citenamefont {Etienne}}]{Saminadayar1997}%
  \BibitemOpen
  \bibfield  {author} {\bibinfo {author} {\bibfnamefont {L.}~\bibnamefont
  {Saminadayar}}, \bibinfo {author} {\bibfnamefont {D.~C.}\ \bibnamefont
  {Glattli}}, \bibinfo {author} {\bibfnamefont {Y.}~\bibnamefont {Jin}}, \ and\
  \bibinfo {author} {\bibfnamefont {B.}~\bibnamefont {Etienne}},\ }\href
  {\doibase 10.1103/PhysRevLett.79.2526} {\bibfield  {journal} {\bibinfo
  {journal} {Phys. Rev. Lett.}\ }\textbf {\bibinfo {volume} {79}},\ \bibinfo
  {pages} {2526} (\bibinfo {year} {1997})}\BibitemShut {NoStop}%
\bibitem [{\citenamefont {{Bakr}}\ \emph {et~al.}(2009)\citenamefont {{Bakr}},
  \citenamefont {{Gillen}}, \citenamefont {{Peng}}, \citenamefont
  {{F{\"o}lling}},\ and\ \citenamefont {{Greiner}}}]{Bakr2009}%
  \BibitemOpen
  \bibfield  {author} {\bibinfo {author} {\bibfnamefont {W.~S.}\ \bibnamefont
  {{Bakr}}}, \bibinfo {author} {\bibfnamefont {J.~I.}\ \bibnamefont
  {{Gillen}}}, \bibinfo {author} {\bibfnamefont {A.}~\bibnamefont {{Peng}}},
  \bibinfo {author} {\bibfnamefont {S.}~\bibnamefont {{F{\"o}lling}}}, \ and\
  \bibinfo {author} {\bibfnamefont {M.}~\bibnamefont {{Greiner}}},\ }\href@noop
  {} {\bibfield  {journal} {\bibinfo  {journal} {\nat}\ }\textbf {\bibinfo
  {volume} {462}},\ \bibinfo {pages} {74} (\bibinfo {year} {2009})}\BibitemShut
  {NoStop}%
\bibitem [{\citenamefont {{Sherson}}\ \emph {et~al.}(2010)\citenamefont
  {{Sherson}}, \citenamefont {{Weitenberg}}, \citenamefont {{Endres}},
  \citenamefont {{Cheneau}}, \citenamefont {{Bloch}},\ and\ \citenamefont
  {{Kuhr}}}]{Sherson2010}%
  \BibitemOpen
  \bibfield  {author} {\bibinfo {author} {\bibfnamefont {J.~F.}\ \bibnamefont
  {{Sherson}}}, \bibinfo {author} {\bibfnamefont {C.}~\bibnamefont
  {{Weitenberg}}}, \bibinfo {author} {\bibfnamefont {M.}~\bibnamefont
  {{Endres}}}, \bibinfo {author} {\bibfnamefont {M.}~\bibnamefont {{Cheneau}}},
  \bibinfo {author} {\bibfnamefont {I.}~\bibnamefont {{Bloch}}}, \ and\
  \bibinfo {author} {\bibfnamefont {S.}~\bibnamefont {{Kuhr}}},\ }\href@noop {}
  {\bibfield  {journal} {\bibinfo  {journal} {\nat}\ }\textbf {\bibinfo
  {volume} {467}},\ \bibinfo {pages} {68} (\bibinfo {year} {2010})}\BibitemShut
  {NoStop}%
\bibitem [{\citenamefont {\"Ottl}\ \emph {et~al.}(2005)\citenamefont {\"Ottl},
  \citenamefont {Ritter}, \citenamefont {K\"ohl},\ and\ \citenamefont
  {Esslinger}}]{Oettl2005}%
  \BibitemOpen
  \bibfield  {author} {\bibinfo {author} {\bibfnamefont {A.}~\bibnamefont
  {\"Ottl}}, \bibinfo {author} {\bibfnamefont {S.}~\bibnamefont {Ritter}},
  \bibinfo {author} {\bibfnamefont {M.}~\bibnamefont {K\"ohl}}, \ and\ \bibinfo
  {author} {\bibfnamefont {T.}~\bibnamefont {Esslinger}},\ }\href {\doibase
  10.1103/PhysRevLett.95.090404} {\bibfield  {journal} {\bibinfo  {journal}
  {Phys. Rev. Lett.}\ }\textbf {\bibinfo {volume} {95}},\ \bibinfo {pages}
  {090404} (\bibinfo {year} {2005})}\BibitemShut {NoStop}%
\bibitem [{\citenamefont {Cherng}\ and\ \citenamefont
  {Demler}(2007)}]{Cherng2007}%
  \BibitemOpen
  \bibfield  {author} {\bibinfo {author} {\bibfnamefont {R.~W.}\ \bibnamefont
  {Cherng}}\ and\ \bibinfo {author} {\bibfnamefont {E.}~\bibnamefont
  {Demler}},\ }\href {http://stacks.iop.org/1367-2630/9/i=1/a=007} {\bibfield
  {journal} {\bibinfo  {journal} {New Journal of Physics}\ }\textbf {\bibinfo
  {volume} {9}},\ \bibinfo {pages} {7} (\bibinfo {year} {2007})}\BibitemShut
  {NoStop}%
\bibitem [{\citenamefont {Belzig}\ \emph {et~al.}(2007)\citenamefont {Belzig},
  \citenamefont {Schroll},\ and\ \citenamefont {Bruder}}]{Belzig2007}%
  \BibitemOpen
  \bibfield  {author} {\bibinfo {author} {\bibfnamefont {W.}~\bibnamefont
  {Belzig}}, \bibinfo {author} {\bibfnamefont {C.}~\bibnamefont {Schroll}}, \
  and\ \bibinfo {author} {\bibfnamefont {C.}~\bibnamefont {Bruder}},\ }\href
  {\doibase 10.1103/PhysRevA.75.063611} {\bibfield  {journal} {\bibinfo
  {journal} {Phys. Rev. A}\ }\textbf {\bibinfo {volume} {75}},\ \bibinfo
  {pages} {063611} (\bibinfo {year} {2007})}\BibitemShut {NoStop}%
\bibitem [{\citenamefont {Braungardt}\ \emph {et~al.}(2008)\citenamefont
  {Braungardt}, \citenamefont {Sen(De)}, \citenamefont {Sen}, \citenamefont
  {Glauber},\ and\ \citenamefont {Lewenstein}}]{Braungardt2008}%
  \BibitemOpen
  \bibfield  {author} {\bibinfo {author} {\bibfnamefont {S.}~\bibnamefont
  {Braungardt}}, \bibinfo {author} {\bibfnamefont {A.}~\bibnamefont {Sen(De)}},
  \bibinfo {author} {\bibfnamefont {U.}~\bibnamefont {Sen}}, \bibinfo {author}
  {\bibfnamefont {R.~J.}\ \bibnamefont {Glauber}}, \ and\ \bibinfo {author}
  {\bibfnamefont {M.}~\bibnamefont {Lewenstein}},\ }\href {\doibase
  10.1103/PhysRevA.78.063613} {\bibfield  {journal} {\bibinfo  {journal} {Phys.
  Rev. A}\ }\textbf {\bibinfo {volume} {78}},\ \bibinfo {pages} {063613}
  (\bibinfo {year} {2008})}\BibitemShut {NoStop}%
\bibitem [{\citenamefont {Lamacraft}(2006)}]{Lamacraft2006}%
  \BibitemOpen
  \bibfield  {author} {\bibinfo {author} {\bibfnamefont {A.}~\bibnamefont
  {Lamacraft}},\ }\href {\doibase 10.1103/PhysRevA.73.011602} {\bibfield
  {journal} {\bibinfo  {journal} {Phys. Rev. A}\ }\textbf {\bibinfo {volume}
  {73}},\ \bibinfo {pages} {011602} (\bibinfo {year} {2006})}\BibitemShut
  {NoStop}%
\bibitem [{\citenamefont {Braungardt}\ \emph {et~al.}(2011)\citenamefont
  {Braungardt}, \citenamefont {Rodr\'{\i}guez}, \citenamefont {Sen(De)},
  \citenamefont {Sen}, \citenamefont {Glauber},\ and\ \citenamefont
  {Lewenstein}}]{Braungardt2011}%
  \BibitemOpen
  \bibfield  {author} {\bibinfo {author} {\bibfnamefont {S.}~\bibnamefont
  {Braungardt}}, \bibinfo {author} {\bibfnamefont {M.}~\bibnamefont
  {Rodr\'{\i}guez}}, \bibinfo {author} {\bibfnamefont {A.}~\bibnamefont
  {Sen(De)}}, \bibinfo {author} {\bibfnamefont {U.}~\bibnamefont {Sen}},
  \bibinfo {author} {\bibfnamefont {R.~J.}\ \bibnamefont {Glauber}}, \ and\
  \bibinfo {author} {\bibfnamefont {M.}~\bibnamefont {Lewenstein}},\ }\href
  {\doibase 10.1103/PhysRevA.83.013601} {\bibfield  {journal} {\bibinfo
  {journal} {Phys. Rev. A}\ }\textbf {\bibinfo {volume} {83}},\ \bibinfo
  {pages} {013601} (\bibinfo {year} {2011})}\BibitemShut {NoStop}%
\bibitem [{\citenamefont {Gring}\ \emph {et~al.}(2012)\citenamefont {Gring},
  \citenamefont {Kuhnert}, \citenamefont {Langen}, \citenamefont {Kitagawa},
  \citenamefont {Rauer}, \citenamefont {Schreitl}, \citenamefont {Mazets},
  \citenamefont {Smith}, \citenamefont {Demler},\ and\ \citenamefont
  {Schmiedmayer}}]{Gring2012}%
  \BibitemOpen
  \bibfield  {author} {\bibinfo {author} {\bibfnamefont {M.}~\bibnamefont
  {Gring}}, \bibinfo {author} {\bibfnamefont {M.}~\bibnamefont {Kuhnert}},
  \bibinfo {author} {\bibfnamefont {T.}~\bibnamefont {Langen}}, \bibinfo
  {author} {\bibfnamefont {T.}~\bibnamefont {Kitagawa}}, \bibinfo {author}
  {\bibfnamefont {B.}~\bibnamefont {Rauer}}, \bibinfo {author} {\bibfnamefont
  {M.}~\bibnamefont {Schreitl}}, \bibinfo {author} {\bibfnamefont
  {I.}~\bibnamefont {Mazets}}, \bibinfo {author} {\bibfnamefont {D.~A.}\
  \bibnamefont {Smith}}, \bibinfo {author} {\bibfnamefont {E.}~\bibnamefont
  {Demler}}, \ and\ \bibinfo {author} {\bibfnamefont {J.}~\bibnamefont
  {Schmiedmayer}},\ }\href@noop {} {\bibfield  {journal} {\bibinfo  {journal}
  {Science}\ }\textbf {\bibinfo {volume} {337}},\ \bibinfo {pages} {1318}
  (\bibinfo {year} {2012})}\BibitemShut {NoStop}%
\bibitem [{\citenamefont {{Haller}}\ \emph {et~al.}(2015)\citenamefont
  {{Haller}}, \citenamefont {{Hudson}}, \citenamefont {{Kelly}}, \citenamefont
  {{Cotta}}, \citenamefont {{Peaudecerf}}, \citenamefont {{Bruce}},\ and\
  \citenamefont {{Kuhr}}}]{Haller2015}%
  \BibitemOpen
  \bibfield  {author} {\bibinfo {author} {\bibfnamefont {E.}~\bibnamefont
  {{Haller}}}, \bibinfo {author} {\bibfnamefont {J.}~\bibnamefont {{Hudson}}},
  \bibinfo {author} {\bibfnamefont {A.}~\bibnamefont {{Kelly}}}, \bibinfo
  {author} {\bibfnamefont {D.~A.}\ \bibnamefont {{Cotta}}}, \bibinfo {author}
  {\bibfnamefont {B.}~\bibnamefont {{Peaudecerf}}}, \bibinfo {author}
  {\bibfnamefont {G.~D.}\ \bibnamefont {{Bruce}}}, \ and\ \bibinfo {author}
  {\bibfnamefont {S.}~\bibnamefont {{Kuhr}}},\ }\href@noop {} {\bibfield
  {journal} {\bibinfo  {journal} {Nature Physics}\ }\textbf {\bibinfo {volume}
  {11}},\ \bibinfo {pages} {738} (\bibinfo {year} {2015})}\BibitemShut
  {NoStop}%
\bibitem [{\citenamefont {Parsons}\ \emph {et~al.}(2016)\citenamefont
  {Parsons}, \citenamefont {Mazurenko}, \citenamefont {Chiu}, \citenamefont
  {Ji}, \citenamefont {Greif},\ and\ \citenamefont {Greiner}}]{Parsons2016}%
  \BibitemOpen
  \bibfield  {author} {\bibinfo {author} {\bibfnamefont {M.~F.}\ \bibnamefont
  {Parsons}}, \bibinfo {author} {\bibfnamefont {A.}~\bibnamefont {Mazurenko}},
  \bibinfo {author} {\bibfnamefont {C.~S.}\ \bibnamefont {Chiu}}, \bibinfo
  {author} {\bibfnamefont {G.}~\bibnamefont {Ji}}, \bibinfo {author}
  {\bibfnamefont {D.}~\bibnamefont {Greif}}, \ and\ \bibinfo {author}
  {\bibfnamefont {M.}~\bibnamefont {Greiner}},\ }\href@noop {} {\bibfield
  {journal} {\bibinfo  {journal} {Science}\ }\textbf {\bibinfo {volume}
  {353}},\ \bibinfo {pages} {1253} (\bibinfo {year} {2016})}\BibitemShut
  {NoStop}%
\bibitem [{\citenamefont {Cheuk}\ \emph {et~al.}(2016)\citenamefont {Cheuk},
  \citenamefont {Nichols}, \citenamefont {Lawrence}, \citenamefont {Okan},
  \citenamefont {Zhang}, \citenamefont {Khatami}, \citenamefont {Trivedi},
  \citenamefont {Paiva}, \citenamefont {Rigol},\ and\ \citenamefont
  {Zwierlein}}]{Cheuk2016}%
  \BibitemOpen
  \bibfield  {author} {\bibinfo {author} {\bibfnamefont {L.~W.}\ \bibnamefont
  {Cheuk}}, \bibinfo {author} {\bibfnamefont {M.~A.}\ \bibnamefont {Nichols}},
  \bibinfo {author} {\bibfnamefont {K.~R.}\ \bibnamefont {Lawrence}}, \bibinfo
  {author} {\bibfnamefont {M.}~\bibnamefont {Okan}}, \bibinfo {author}
  {\bibfnamefont {H.}~\bibnamefont {Zhang}}, \bibinfo {author} {\bibfnamefont
  {E.}~\bibnamefont {Khatami}}, \bibinfo {author} {\bibfnamefont
  {N.}~\bibnamefont {Trivedi}}, \bibinfo {author} {\bibfnamefont
  {T.}~\bibnamefont {Paiva}}, \bibinfo {author} {\bibfnamefont
  {M.}~\bibnamefont {Rigol}}, \ and\ \bibinfo {author} {\bibfnamefont {M.~W.}\
  \bibnamefont {Zwierlein}},\ }\href@noop {} {\bibfield  {journal} {\bibinfo
  {journal} {Science}\ }\textbf {\bibinfo {volume} {353}},\ \bibinfo {pages}
  {1260} (\bibinfo {year} {2016})}\BibitemShut {NoStop}%
\bibitem [{\citenamefont {Drewes}\ \emph {et~al.}(2016)\citenamefont {Drewes},
  \citenamefont {Cocchi}, \citenamefont {Miller}, \citenamefont {Chan},
  \citenamefont {Pertot}, \citenamefont {Brennecke},\ and\ \citenamefont
  {K\"ohl}}]{Drewes2016}%
  \BibitemOpen
  \bibfield  {author} {\bibinfo {author} {\bibfnamefont {J.~H.}\ \bibnamefont
  {Drewes}}, \bibinfo {author} {\bibfnamefont {E.}~\bibnamefont {Cocchi}},
  \bibinfo {author} {\bibfnamefont {L.~A.}\ \bibnamefont {Miller}}, \bibinfo
  {author} {\bibfnamefont {C.~F.}\ \bibnamefont {Chan}}, \bibinfo {author}
  {\bibfnamefont {D.}~\bibnamefont {Pertot}}, \bibinfo {author} {\bibfnamefont
  {F.}~\bibnamefont {Brennecke}}, \ and\ \bibinfo {author} {\bibfnamefont
  {M.}~\bibnamefont {K\"ohl}},\ }\href {\doibase
  10.1103/PhysRevLett.117.135301} {\bibfield  {journal} {\bibinfo  {journal}
  {Phys. Rev. Lett.}\ }\textbf {\bibinfo {volume} {117}},\ \bibinfo {pages}
  {135301} (\bibinfo {year} {2016})}\BibitemShut {NoStop}%
\bibitem [{\citenamefont {{Mazurenko}}\ \emph {et~al.}(2017)\citenamefont
  {{Mazurenko}}, \citenamefont {{Chiu}}, \citenamefont {{Ji}}, \citenamefont
  {{Parsons}}, \citenamefont {{Kan{\'a}sz-Nagy}}, \citenamefont {{Schmidt}},
  \citenamefont {{Grusdt}}, \citenamefont {{Demler}}, \citenamefont {{Greif}},\
  and\ \citenamefont {{Greiner}}}]{Mazurenko2017}%
  \BibitemOpen
  \bibfield  {author} {\bibinfo {author} {\bibfnamefont {A.}~\bibnamefont
  {{Mazurenko}}}, \bibinfo {author} {\bibfnamefont {C.~S.}\ \bibnamefont
  {{Chiu}}}, \bibinfo {author} {\bibfnamefont {G.}~\bibnamefont {{Ji}}},
  \bibinfo {author} {\bibfnamefont {M.~F.}\ \bibnamefont {{Parsons}}}, \bibinfo
  {author} {\bibfnamefont {M.}~\bibnamefont {{Kan{\'a}sz-Nagy}}}, \bibinfo
  {author} {\bibfnamefont {R.}~\bibnamefont {{Schmidt}}}, \bibinfo {author}
  {\bibfnamefont {F.}~\bibnamefont {{Grusdt}}}, \bibinfo {author}
  {\bibfnamefont {E.}~\bibnamefont {{Demler}}}, \bibinfo {author}
  {\bibfnamefont {D.}~\bibnamefont {{Greif}}}, \ and\ \bibinfo {author}
  {\bibfnamefont {M.}~\bibnamefont {{Greiner}}},\ }\href@noop {} {\bibfield
  {journal} {\bibinfo  {journal} {Nature}\ }\textbf {\bibinfo {volume} {545}},\
  \bibinfo {pages} {462} (\bibinfo {year} {2017})}\BibitemShut {NoStop}%
\bibitem [{\citenamefont {{Hilker}}\ \emph {et~al.}(2017)\citenamefont
  {{Hilker}}, \citenamefont {{Salomon}}, \citenamefont {{Grusdt}},
  \citenamefont {{Omran}}, \citenamefont {{Boll}}, \citenamefont {{Demler}},
  \citenamefont {{Bloch}},\ and\ \citenamefont {{Gross}}}]{Hilker2017}%
  \BibitemOpen
  \bibfield  {author} {\bibinfo {author} {\bibfnamefont {T.~A.}\ \bibnamefont
  {{Hilker}}}, \bibinfo {author} {\bibfnamefont {G.}~\bibnamefont {{Salomon}}},
  \bibinfo {author} {\bibfnamefont {F.}~\bibnamefont {{Grusdt}}}, \bibinfo
  {author} {\bibfnamefont {A.}~\bibnamefont {{Omran}}}, \bibinfo {author}
  {\bibfnamefont {M.}~\bibnamefont {{Boll}}}, \bibinfo {author} {\bibfnamefont
  {E.}~\bibnamefont {{Demler}}}, \bibinfo {author} {\bibfnamefont
  {I.}~\bibnamefont {{Bloch}}}, \ and\ \bibinfo {author} {\bibfnamefont
  {C.}~\bibnamefont {{Gross}}},\ }\href@noop {} {\bibfield  {journal} {\bibinfo
   {journal} {ArXiv e-prints}\ } (\bibinfo {year} {2017})},\ \Eprint
  {http://arxiv.org/abs/1702.00642} {arXiv:1702.00642} \BibitemShut {NoStop}%
\bibitem [{\citenamefont {{Mitra}}\ \emph {et~al.}(2017)\citenamefont
  {{Mitra}}, \citenamefont {{Brown}}, \citenamefont {{Guardado-Sanchez}},
  \citenamefont {{Kondov}}, \citenamefont {{Devakul}}, \citenamefont {{Huse}},
  \citenamefont {{Schauss}},\ and\ \citenamefont {{Bakr}}}]{Mitra2017}%
  \BibitemOpen
  \bibfield  {author} {\bibinfo {author} {\bibfnamefont {D.}~\bibnamefont
  {{Mitra}}}, \bibinfo {author} {\bibfnamefont {P.~T.}\ \bibnamefont
  {{Brown}}}, \bibinfo {author} {\bibfnamefont {E.}~\bibnamefont
  {{Guardado-Sanchez}}}, \bibinfo {author} {\bibfnamefont {S.~S.}\ \bibnamefont
  {{Kondov}}}, \bibinfo {author} {\bibfnamefont {T.}~\bibnamefont {{Devakul}}},
  \bibinfo {author} {\bibfnamefont {D.~A.}\ \bibnamefont {{Huse}}}, \bibinfo
  {author} {\bibfnamefont {P.}~\bibnamefont {{Schauss}}}, \ and\ \bibinfo
  {author} {\bibfnamefont {W.~S.}\ \bibnamefont {{Bakr}}},\ }\href@noop {}
  {\bibfield  {journal} {\bibinfo  {journal} {ArXiv e-prints}\ } (\bibinfo
  {year} {2017})},\ \Eprint {http://arxiv.org/abs/1705.02039}
  {arXiv:1705.02039} \BibitemShut {NoStop}%
\bibitem [{Ass()}]{Assaad2008}%
  \BibitemOpen
  \href@noop {} {}\bibinfo {note} {For reviews consult, e.g., R. R. dos Santos,
  Introduction to Quantum Monte Carlo Simulations for Fermionic Systems. Braz.
  J. Phys. {\bf 33}, 36 (2003); F. F. Assaad and H. G. Evertz, Worldline and
  Determinantal Quantum Monte Carlo Methods for Spins, Phonons and Electrons.
  Lecture Notes in Physics {\bf 739}, 277 (2008).}\BibitemShut {Stop}%
\bibitem [{\citenamefont {Grover}(2013)}]{Grover2013}%
  \BibitemOpen
  \bibfield  {author} {\bibinfo {author} {\bibfnamefont {T.}~\bibnamefont
  {Grover}},\ }\href {\doibase 10.1103/PhysRevLett.111.130402} {\bibfield
  {journal} {\bibinfo  {journal} {Phys. Rev. Lett.}\ }\textbf {\bibinfo
  {volume} {111}},\ \bibinfo {pages} {130402} (\bibinfo {year}
  {2013})}\BibitemShut {NoStop}%
\bibitem [{\citenamefont {Cheong}\ and\ \citenamefont
  {Henley}(2004)}]{Cheong2004}%
  \BibitemOpen
  \bibfield  {author} {\bibinfo {author} {\bibfnamefont {S.-A.}\ \bibnamefont
  {Cheong}}\ and\ \bibinfo {author} {\bibfnamefont {C.~L.}\ \bibnamefont
  {Henley}},\ }\href {\doibase 10.1103/PhysRevB.69.075111} {\bibfield
  {journal} {\bibinfo  {journal} {Phys. Rev. B}\ }\textbf {\bibinfo {volume}
  {69}},\ \bibinfo {pages} {075111} (\bibinfo {year} {2004})}\BibitemShut
  {NoStop}%
\bibitem [{\citenamefont {Peschel}(2003)}]{Peschel2003}%
  \BibitemOpen
  \bibfield  {author} {\bibinfo {author} {\bibfnamefont {I.}~\bibnamefont
  {Peschel}},\ }\href {http://stacks.iop.org/0305-4470/36/i=14/a=101}
  {\bibfield  {journal} {\bibinfo  {journal} {Journal of Physics A:
  Mathematical and General}\ }\textbf {\bibinfo {volume} {36}},\ \bibinfo
  {pages} {L205} (\bibinfo {year} {2003})}\BibitemShut {NoStop}%
\bibitem [{\citenamefont {Chung}\ and\ \citenamefont
  {Peschel}(2001)}]{Chung2001}%
  \BibitemOpen
  \bibfield  {author} {\bibinfo {author} {\bibfnamefont {M.-C.}\ \bibnamefont
  {Chung}}\ and\ \bibinfo {author} {\bibfnamefont {I.}~\bibnamefont
  {Peschel}},\ }\href {\doibase 10.1103/PhysRevB.64.064412} {\bibfield
  {journal} {\bibinfo  {journal} {Phys. Rev. B}\ }\textbf {\bibinfo {volume}
  {64}},\ \bibinfo {pages} {064412} (\bibinfo {year} {2001})}\BibitemShut
  {NoStop}%
\bibitem [{\citenamefont {Micnas}\ \emph {et~al.}(1990)\citenamefont {Micnas},
  \citenamefont {Ranninger},\ and\ \citenamefont {Robaszkiewicz}}]{Micnas1990}%
  \BibitemOpen
  \bibfield  {author} {\bibinfo {author} {\bibfnamefont {R.}~\bibnamefont
  {Micnas}}, \bibinfo {author} {\bibfnamefont {J.}~\bibnamefont {Ranninger}}, \
  and\ \bibinfo {author} {\bibfnamefont {S.}~\bibnamefont {Robaszkiewicz}},\
  }\href {\doibase 10.1103/RevModPhys.62.113} {\bibfield  {journal} {\bibinfo
  {journal} {Rev. Mod. Phys.}\ }\textbf {\bibinfo {volume} {62}},\ \bibinfo
  {pages} {113} (\bibinfo {year} {1990})}\BibitemShut {NoStop}%
\bibitem [{\citenamefont {Paiva}\ \emph {et~al.}(2004)\citenamefont {Paiva},
  \citenamefont {dos Santos}, \citenamefont {Scalettar},\ and\ \citenamefont
  {Denteneer}}]{Paiva2004}%
  \BibitemOpen
  \bibfield  {author} {\bibinfo {author} {\bibfnamefont {T.}~\bibnamefont
  {Paiva}}, \bibinfo {author} {\bibfnamefont {R.~R.}\ \bibnamefont {dos
  Santos}}, \bibinfo {author} {\bibfnamefont {R.~T.}\ \bibnamefont
  {Scalettar}}, \ and\ \bibinfo {author} {\bibfnamefont {P.~J.~H.}\
  \bibnamefont {Denteneer}},\ }\href {\doibase 10.1103/PhysRevB.69.184501}
  {\bibfield  {journal} {\bibinfo  {journal} {Phys. Rev. B}\ }\textbf {\bibinfo
  {volume} {69}},\ \bibinfo {pages} {184501} (\bibinfo {year}
  {2004})}\BibitemShut {NoStop}%
\bibitem [{\citenamefont {Paiva}\ \emph {et~al.}(2010)\citenamefont {Paiva},
  \citenamefont {Scalettar}, \citenamefont {Randeria},\ and\ \citenamefont
  {Trivedi}}]{Paiva2010}%
  \BibitemOpen
  \bibfield  {author} {\bibinfo {author} {\bibfnamefont {T.}~\bibnamefont
  {Paiva}}, \bibinfo {author} {\bibfnamefont {R.}~\bibnamefont {Scalettar}},
  \bibinfo {author} {\bibfnamefont {M.}~\bibnamefont {Randeria}}, \ and\
  \bibinfo {author} {\bibfnamefont {N.}~\bibnamefont {Trivedi}},\ }\href
  {\doibase 10.1103/PhysRevLett.104.066406} {\bibfield  {journal} {\bibinfo
  {journal} {Phys. Rev. Lett.}\ }\textbf {\bibinfo {volume} {104}},\ \bibinfo
  {pages} {066406} (\bibinfo {year} {2010})}\BibitemShut {NoStop}%
\bibitem [{\citenamefont {Eagles}(1969)}]{Eagles1969}%
  \BibitemOpen
  \bibfield  {author} {\bibinfo {author} {\bibfnamefont {D.~M.}\ \bibnamefont
  {Eagles}},\ }\href {\doibase 10.1103/PhysRev.186.456} {\bibfield  {journal}
  {\bibinfo  {journal} {Phys. Rev.}\ }\textbf {\bibinfo {volume} {186}},\
  \bibinfo {pages} {456} (\bibinfo {year} {1969})}\BibitemShut {NoStop}%
\bibitem [{Leg()}]{Leggett1980}%
  \BibitemOpen
  \href@noop {} {}\bibinfo {note} {A. J. Leggett, in \emph{Modern Trends in the
  Theory of Condensed Matter}, edited by A. Pedalski and J. Przystawa
  (Springer, Berlin, 1980).}\BibitemShut {Stop}%
\bibitem [{\citenamefont {Nozi{\`e}res}\ and\ \citenamefont
  {Schmitt-Rink}(1985)}]{Nozieres1985}%
  \BibitemOpen
  \bibfield  {author} {\bibinfo {author} {\bibfnamefont {P.}~\bibnamefont
  {Nozi{\`e}res}}\ and\ \bibinfo {author} {\bibfnamefont {S.}~\bibnamefont
  {Schmitt-Rink}},\ }\href {\doibase 10.1007/BF00683774} {\bibfield  {journal}
  {\bibinfo  {journal} {Journal of Low Temperature Physics}\ }\textbf {\bibinfo
  {volume} {59}},\ \bibinfo {pages} {195} (\bibinfo {year} {1985})}\BibitemShut
  {NoStop}%
\bibitem [{\citenamefont {Landau}\ and\ \citenamefont
  {Lifshitz}(1987)}]{LandauLifshitz}%
  \BibitemOpen
  \bibfield  {author} {\bibinfo {author} {\bibfnamefont {L.~D.}\ \bibnamefont
  {Landau}}\ and\ \bibinfo {author} {\bibfnamefont {E.~M.}\ \bibnamefont
  {Lifshitz}},\ }\href@noop {} {\emph {\bibinfo {title} {Quantum Mechanics}}},\
  \bibinfo {edition} {3rd}\ ed.\ (\bibinfo  {publisher} {Pergamon, Oxford},\
  \bibinfo {year} {1987})\ \bibinfo {note} {sec. 133}\BibitemShut {NoStop}%
\bibitem [{\citenamefont {Randeria}\ \emph {et~al.}(1990)\citenamefont
  {Randeria}, \citenamefont {Duan},\ and\ \citenamefont
  {Shieh}}]{Randeria1990}%
  \BibitemOpen
  \bibfield  {author} {\bibinfo {author} {\bibfnamefont {M.}~\bibnamefont
  {Randeria}}, \bibinfo {author} {\bibfnamefont {J.-M.}\ \bibnamefont {Duan}},
  \ and\ \bibinfo {author} {\bibfnamefont {L.-Y.}\ \bibnamefont {Shieh}},\
  }\href {\doibase 10.1103/PhysRevB.41.327} {\bibfield  {journal} {\bibinfo
  {journal} {Phys. Rev. B}\ }\textbf {\bibinfo {volume} {41}},\ \bibinfo
  {pages} {327} (\bibinfo {year} {1990})}\BibitemShut {NoStop}%
\bibitem [{\citenamefont {MacDonald}\ \emph {et~al.}(1988)\citenamefont
  {MacDonald}, \citenamefont {Girvin},\ and\ \citenamefont
  {Yoshioka}}]{MacDonald1988}%
  \BibitemOpen
  \bibfield  {author} {\bibinfo {author} {\bibfnamefont {A.~H.}\ \bibnamefont
  {MacDonald}}, \bibinfo {author} {\bibfnamefont {S.~M.}\ \bibnamefont
  {Girvin}}, \ and\ \bibinfo {author} {\bibfnamefont {D.}~\bibnamefont
  {Yoshioka}},\ }\href {\doibase 10.1103/PhysRevB.37.9753} {\bibfield
  {journal} {\bibinfo  {journal} {Phys. Rev. B}\ }\textbf {\bibinfo {volume}
  {37}},\ \bibinfo {pages} {9753} (\bibinfo {year} {1988})}\BibitemShut
  {NoStop}%
\bibitem [{\citenamefont {Takahashi}(1977)}]{Takahashi1977}%
  \BibitemOpen
  \bibfield  {author} {\bibinfo {author} {\bibfnamefont {M.}~\bibnamefont
  {Takahashi}},\ }\href {http://stacks.iop.org/0022-3719/10/i=8/a=031}
  {\bibfield  {journal} {\bibinfo  {journal} {Journal of Physics C: Solid State
  Physics}\ }\textbf {\bibinfo {volume} {10}},\ \bibinfo {pages} {1289}
  (\bibinfo {year} {1977})}\BibitemShut {NoStop}%
\bibitem [{\citenamefont {Delannoy}\ \emph {et~al.}(2005)\citenamefont
  {Delannoy}, \citenamefont {Gingras}, \citenamefont {Holdsworth},\ and\
  \citenamefont {Tremblay}}]{Delannoy2005}%
  \BibitemOpen
  \bibfield  {author} {\bibinfo {author} {\bibfnamefont {J.-Y.~P.}\
  \bibnamefont {Delannoy}}, \bibinfo {author} {\bibfnamefont {M.~J.~P.}\
  \bibnamefont {Gingras}}, \bibinfo {author} {\bibfnamefont {P.~C.~W.}\
  \bibnamefont {Holdsworth}}, \ and\ \bibinfo {author} {\bibfnamefont
  {A.-M.~S.}\ \bibnamefont {Tremblay}},\ }\href {\doibase
  10.1103/PhysRevB.72.115114} {\bibfield  {journal} {\bibinfo  {journal} {Phys.
  Rev. B}\ }\textbf {\bibinfo {volume} {72}},\ \bibinfo {pages} {115114}
  (\bibinfo {year} {2005})}\BibitemShut {NoStop}%
\bibitem [{\citenamefont {{Lovas}}\ \emph {et~al.}(2017)\citenamefont
  {{Lovas}}, \citenamefont {{D{\'o}ra}}, \citenamefont {{Demler}},\ and\
  \citenamefont {{Zar{\'a}nd}}}]{Lovas2017}%
  \BibitemOpen
  \bibfield  {author} {\bibinfo {author} {\bibfnamefont {I.}~\bibnamefont
  {{Lovas}}}, \bibinfo {author} {\bibfnamefont {B.}~\bibnamefont {{D{\'o}ra}}},
  \bibinfo {author} {\bibfnamefont {E.}~\bibnamefont {{Demler}}}, \ and\
  \bibinfo {author} {\bibfnamefont {G.}~\bibnamefont {{Zar{\'a}nd}}},\
  }\href@noop {} {\bibfield  {journal} {\bibinfo  {journal} {\pra}\ }\textbf
  {\bibinfo {volume} {95}},\ \bibinfo {pages} {053621} (\bibinfo {year}
  {2017})}\BibitemShut {NoStop}%
\bibitem [{\citenamefont {Meng}\ \emph {et~al.}(2010)\citenamefont {Meng},
  \citenamefont {Lang}, \citenamefont {Wessel}, \citenamefont {Assaad},\ and\
  \citenamefont {Muramatsu}}]{Meng2010}%
  \BibitemOpen
  \bibfield  {author} {\bibinfo {author} {\bibfnamefont {Z.~Y.}\ \bibnamefont
  {Meng}}, \bibinfo {author} {\bibfnamefont {T.~C.}\ \bibnamefont {Lang}},
  \bibinfo {author} {\bibfnamefont {S.}~\bibnamefont {Wessel}}, \bibinfo
  {author} {\bibfnamefont {F.~F.}\ \bibnamefont {Assaad}}, \ and\ \bibinfo
  {author} {\bibfnamefont {A.}~\bibnamefont {Muramatsu}},\ }\href@noop {}
  {\bibfield  {journal} {\bibinfo  {journal} {Nature}\ }\textbf {\bibinfo
  {volume} {464}},\ \bibinfo {pages} {847} (\bibinfo {year}
  {2010})}\BibitemShut {NoStop}%
\bibitem [{\citenamefont {{Sorella}}\ \emph {et~al.}(2012)\citenamefont
  {{Sorella}}, \citenamefont {{Otsuka}},\ and\ \citenamefont
  {{Yunoki}}}]{Sorella2012}%
  \BibitemOpen
  \bibfield  {author} {\bibinfo {author} {\bibfnamefont {S.}~\bibnamefont
  {{Sorella}}}, \bibinfo {author} {\bibfnamefont {Y.}~\bibnamefont {{Otsuka}}},
  \ and\ \bibinfo {author} {\bibfnamefont {S.}~\bibnamefont {{Yunoki}}},\
  }\href@noop {} {\bibfield  {journal} {\bibinfo  {journal} {Scientific
  Reports}\ }\textbf {\bibinfo {volume} {2}},\ \bibinfo {pages} {992} (\bibinfo
  {year} {2012})}\BibitemShut {NoStop}%
\end{thebibliography}

\begin{thebibliography}{13}%
\makeatletter
\providecommand \@ifxundefined [1]{%
 \@ifx{#1\undefined}
}%
\providecommand \@ifnum [1]{%
 \ifnum #1\expandafter \@firstoftwo
 \else \expandafter \@secondoftwo
 \fi
}%
\providecommand \@ifx [1]{%
 \ifx #1\expandafter \@firstoftwo
 \else \expandafter \@secondoftwo
 \fi
}%
\providecommand \natexlab [1]{#1}%
\providecommand \enquote  [1]{``#1''}%
\providecommand \bibnamefont  [1]{#1}%
\providecommand \bibfnamefont [1]{#1}%
\providecommand \citenamefont [1]{#1}%
\providecommand \href@noop [0]{\@secondoftwo}%
\providecommand \href [0]{\begingroup \@sanitize@url \@href}%
\providecommand \@href[1]{\@@startlink{#1}\@@href}%
\providecommand \@@href[1]{\endgroup#1\@@endlink}%
\providecommand \@sanitize@url [0]{\catcode `\\12\catcode `\$12\catcode
  `\&12\catcode `\#12\catcode `\^12\catcode `\_12\catcode `\%12\relax}%
\providecommand \@@startlink[1]{}%
\providecommand \@@endlink[0]{}%
\providecommand \url  [0]{\begingroup\@sanitize@url \@url }%
\providecommand \@url [1]{\endgroup\@href {#1}{\urlprefix }}%
\providecommand \urlprefix  [0]{URL }%
\providecommand \Eprint [0]{\href }%
\providecommand \doibase [0]{http://dx.doi.org/}%
\providecommand \selectlanguage [0]{\@gobble}%
\providecommand \bibinfo  [0]{\@secondoftwo}%
\providecommand \bibfield  [0]{\@secondoftwo}%
\providecommand \translation [1]{[#1]}%
\providecommand \BibitemOpen [0]{}%
\providecommand \bibitemStop [0]{}%
\providecommand \bibitemNoStop [0]{.\EOS\space}%
\providecommand \EOS [0]{\spacefactor3000\relax}%
\providecommand \BibitemShut  [1]{\csname bibitem#1\endcsname}%
\let\auto@bib@innerbib\@empty
\bibitem [{\citenamefont {Al-Mohy}\ and\ \citenamefont
  {Higham}(2012)}]{Al-Mohy2012}%
  \BibitemOpen
  \bibfield  {author} {\bibinfo {author} {\bibfnamefont {A.~H.}\ \bibnamefont
  {Al-Mohy}}\ and\ \bibinfo {author} {\bibfnamefont {N.~J.}\ \bibnamefont
  {Higham}},\ }\href@noop {} {\bibfield  {journal} {\bibinfo  {journal} {SIAM
  J. Sci. Comput.}\ }\textbf {\bibinfo {volume} {34}},\ \bibinfo {pages} {C152}
  (\bibinfo {year} {2012})}\BibitemShut {NoStop}%
\bibitem [{\citenamefont {Schrieffer}(1964)}]{Schrieffer_book}%
  \BibitemOpen
  \bibfield  {author} {\bibinfo {author} {\bibfnamefont {J.~R.}\ \bibnamefont
  {Schrieffer}},\ }\href@noop {} {\emph {\bibinfo {title} {Theory of
  Superconductivity}}},\ Frontiers in Physics, Reading, MA\ (\bibinfo {year}
  {1964})\BibitemShut {NoStop}%
\bibitem [{Leg()}]{Leggett1980_SM}%
  \BibitemOpen
  \href@noop {} {}\bibinfo {note} {A. J. Leggett, in \emph{Modern Trends in the
  Theory of Condensed Matter}, edited by A. Pedalski and J. Przystawa
  (Springer, Berlin, 1980).}\BibitemShut {Stop}%
\bibitem [{\citenamefont {Cherng}\ and\ \citenamefont
  {Demler}(2007)}]{Cherng2007_SM}%
  \BibitemOpen
  \bibfield  {author} {\bibinfo {author} {\bibfnamefont {R.~W.}\ \bibnamefont
  {Cherng}}\ and\ \bibinfo {author} {\bibfnamefont {E.}~\bibnamefont
  {Demler}},\ }\href {http://stacks.iop.org/1367-2630/9/i=1/a=007} {\bibfield
  {journal} {\bibinfo  {journal} {New Journal of Physics}\ }\textbf {\bibinfo
  {volume} {9}},\ \bibinfo {pages} {7} (\bibinfo {year} {2007})}\BibitemShut
  {NoStop}%
\bibitem [{\citenamefont {Micnas}\ \emph {et~al.}(1990)\citenamefont {Micnas},
  \citenamefont {Ranninger},\ and\ \citenamefont {Robaszkiewicz}}]{Micnas1990_SM}%
  \BibitemOpen
  \bibfield  {author} {\bibinfo {author} {\bibfnamefont {R.}~\bibnamefont
  {Micnas}}, \bibinfo {author} {\bibfnamefont {J.}~\bibnamefont {Ranninger}}, \
  and\ \bibinfo {author} {\bibfnamefont {S.}~\bibnamefont {Robaszkiewicz}},\
  }\href {\doibase 10.1103/RevModPhys.62.113} {\bibfield  {journal} {\bibinfo
  {journal} {Rev. Mod. Phys.}\ }\textbf {\bibinfo {volume} {62}},\ \bibinfo
  {pages} {113} (\bibinfo {year} {1990})}\BibitemShut {NoStop}%
\bibitem [{\citenamefont {Belkhir}\ and\ \citenamefont
  {Randeria}(1992)}]{Belkhir1992_SM}%
  \BibitemOpen
  \bibfield  {author} {\bibinfo {author} {\bibfnamefont {L.}~\bibnamefont
  {Belkhir}}\ and\ \bibinfo {author} {\bibfnamefont {M.}~\bibnamefont
  {Randeria}},\ }\href {\doibase 10.1103/PhysRevB.45.5087} {\bibfield
  {journal} {\bibinfo  {journal} {Phys. Rev. B}\ }\textbf {\bibinfo {volume}
  {45}},\ \bibinfo {pages} {5087} (\bibinfo {year} {1992})}\BibitemShut
  {NoStop}%
\bibitem [{\citenamefont {Eagles}(1969)}]{Eagles1969_SM}%
  \BibitemOpen
  \bibfield  {author} {\bibinfo {author} {\bibfnamefont {D.~M.}\ \bibnamefont
  {Eagles}},\ }\href {\doibase 10.1103/PhysRev.186.456} {\bibfield  {journal}
  {\bibinfo  {journal} {Phys. Rev.}\ }\textbf {\bibinfo {volume} {186}},\
  \bibinfo {pages} {456} (\bibinfo {year} {1969})}\BibitemShut {NoStop}%
\bibitem [{\citenamefont {Nozi{\`e}res}\ and\ \citenamefont
  {Schmitt-Rink}(1985)}]{Nozieres1985_SM}%
  \BibitemOpen
  \bibfield  {author} {\bibinfo {author} {\bibfnamefont {P.}~\bibnamefont
  {Nozi{\`e}res}}\ and\ \bibinfo {author} {\bibfnamefont {S.}~\bibnamefont
  {Schmitt-Rink}},\ }\href {\doibase 10.1007/BF00683774} {\bibfield  {journal}
  {\bibinfo  {journal} {Journal of Low Temperature Physics}\ }\textbf {\bibinfo
  {volume} {59}},\ \bibinfo {pages} {195} (\bibinfo {year} {1985})}\BibitemShut
  {NoStop}%
\bibitem [{\citenamefont {Belzig}\ \emph {et~al.}(2007)\citenamefont {Belzig},
  \citenamefont {Schroll},\ and\ \citenamefont {Bruder}}]{Belzig2007_SM}%
  \BibitemOpen
  \bibfield  {author} {\bibinfo {author} {\bibfnamefont {W.}~\bibnamefont
  {Belzig}}, \bibinfo {author} {\bibfnamefont {C.}~\bibnamefont {Schroll}}, \
  and\ \bibinfo {author} {\bibfnamefont {C.}~\bibnamefont {Bruder}},\ }\href
  {\doibase 10.1103/PhysRevA.75.063611} {\bibfield  {journal} {\bibinfo
  {journal} {Phys. Rev. A}\ }\textbf {\bibinfo {volume} {75}},\ \bibinfo
  {pages} {063611} (\bibinfo {year} {2007})}\BibitemShut {NoStop}%
\bibitem [{\citenamefont {Delannoy}\ \emph {et~al.}(2005)\citenamefont
  {Delannoy}, \citenamefont {Gingras}, \citenamefont {Holdsworth},\ and\
  \citenamefont {Tremblay}}]{Delannoy2005_SM}%
  \BibitemOpen
  \bibfield  {author} {\bibinfo {author} {\bibfnamefont {J.-Y.~P.}\
  \bibnamefont {Delannoy}}, \bibinfo {author} {\bibfnamefont {M.~J.~P.}\
  \bibnamefont {Gingras}}, \bibinfo {author} {\bibfnamefont {P.~C.~W.}\
  \bibnamefont {Holdsworth}}, \ and\ \bibinfo {author} {\bibfnamefont
  {A.-M.~S.}\ \bibnamefont {Tremblay}},\ }\href {\doibase
  10.1103/PhysRevB.72.115114} {\bibfield  {journal} {\bibinfo  {journal} {Phys.
  Rev. B}\ }\textbf {\bibinfo {volume} {72}},\ \bibinfo {pages} {115114}
  (\bibinfo {year} {2005})}\BibitemShut {NoStop}%
\bibitem [{\citenamefont {{Mazurenko}}\ \emph {et~al.}(2017)\citenamefont
  {{Mazurenko}}, \citenamefont {{Chiu}}, \citenamefont {{Ji}}, \citenamefont
  {{Parsons}}, \citenamefont {{Kan{\'a}sz-Nagy}}, \citenamefont {{Schmidt}},
  \citenamefont {{Grusdt}}, \citenamefont {{Demler}}, \citenamefont {{Greif}},\
  and\ \citenamefont {{Greiner}}}]{Mazurenko2017_SM}%
  \BibitemOpen
  \bibfield  {author} {\bibinfo {author} {\bibfnamefont {A.}~\bibnamefont
  {{Mazurenko}}}, \bibinfo {author} {\bibfnamefont {C.~S.}\ \bibnamefont
  {{Chiu}}}, \bibinfo {author} {\bibfnamefont {G.}~\bibnamefont {{Ji}}},
  \bibinfo {author} {\bibfnamefont {M.~F.}\ \bibnamefont {{Parsons}}}, \bibinfo
  {author} {\bibfnamefont {M.}~\bibnamefont {{Kan{\'a}sz-Nagy}}}, \bibinfo
  {author} {\bibfnamefont {R.}~\bibnamefont {{Schmidt}}}, \bibinfo {author}
  {\bibfnamefont {F.}~\bibnamefont {{Grusdt}}}, \bibinfo {author}
  {\bibfnamefont {E.}~\bibnamefont {{Demler}}}, \bibinfo {author}
  {\bibfnamefont {D.}~\bibnamefont {{Greif}}}, \ and\ \bibinfo {author}
  {\bibfnamefont {M.}~\bibnamefont {{Greiner}}},\ }\href@noop {} {\bibfield
  {journal} {\bibinfo  {journal} {Nature}\ }\textbf {\bibinfo {volume} {545}},\
  \bibinfo {pages} {462} (\bibinfo {year} {2017})}\BibitemShut {NoStop}%
\bibitem [{\citenamefont {Sandvik}(1999)}]{Sandvik1999_SM}%
  \BibitemOpen
  \bibfield  {author} {\bibinfo {author} {\bibfnamefont {A.~W.}\ \bibnamefont
  {Sandvik}},\ }\href {\doibase 10.1103/PhysRevB.59.R14157} {\bibfield
  {journal} {\bibinfo  {journal} {Phys. Rev. B}\ }\textbf {\bibinfo {volume}
  {59}},\ \bibinfo {pages} {R14157} (\bibinfo {year} {1999})}\BibitemShut
  {NoStop}%
\bibitem [{\citenamefont {Sandvik}(1992)}]{Sandvik1992_SM}%
  \BibitemOpen
  \bibfield  {author} {\bibinfo {author} {\bibfnamefont {A.~W.}\ \bibnamefont
  {Sandvik}},\ }\href {http://stacks.iop.org/0305-4470/25/i=13/a=017}
  {\bibfield  {journal} {\bibinfo  {journal} {Journal of Physics A:
  Mathematical and General}\ }\textbf {\bibinfo {volume} {25}},\ \bibinfo
  {pages} {3667} (\bibinfo {year} {1992})}\BibitemShut {NoStop}%
\end{thebibliography}
\end{document}